\pdfoutput=1
\documentclass[11pt,letterpaper]{article}
\usepackage[letterpaper,margin=1in]{geometry}
\usepackage{amsmath,amssymb,amsthm,mathtools}
\usepackage{booktabs,array}
\usepackage[hidelinks]{hyperref}
\usepackage[capitalise,nameinlink]{cleveref}
\usepackage{xcolor}
\usepackage{tikz}
\usetikzlibrary{arrows.meta,decorations.pathreplacing}
\usepackage{pgfplots}
\pgfplotsset{compat=1.18}

\newtheorem{theorem}{Theorem}[section]
\newtheorem{lemma}[theorem]{Lemma}
\newtheorem{proposition}[theorem]{Proposition}

\theoremstyle{remark}
\newtheorem{remark}[theorem]{Remark}

\newcommand{\MC}[1]{\textup{\textsc{Max-$#1$-Cut}}}
\newcommand{\lin}{\textup{\textsc{3Lin}}(\mathbb{Z}_3)}
\newcommand{\Z}{\mathbb{Z}}
\newcommand{\R}{\mathbb{R}}
\newcommand{\C}{\mathbb{C}}
\newcommand{\eps}{\varepsilon}
\newcommand{\Y}{\mathcal{Y}}
\newcommand{\sat}{\mathrm{sat}}
\newcommand{\cut}{\mathrm{cut}}
\newcommand{\val}{\mathrm{val}}
\newcommand{\col}{\mathrm{col}}
\newcommand{\can}{\mathrm{can}}
\newcommand{\Sym}{\mathrm{Sym}}
\newcommand{\yc}{y^{\circ}}

\title{Improved NP-Hardness of Approximation for \MC{3}\\ and for \MC{k} with Small $k$}
\author{Ziang Ni\\[2pt] {\normalsize Delft University of Technology}\\ {\normalsize\texttt{ziangni@tudelft.nl}}}
\date{}

\begin{document}
\maketitle
\thispagestyle{empty}

\begin{abstract}
We prove that \MC{3} is NP-hard to approximate within a factor of $27/29+\eps\approx 0.931$. This
improves the factor $16/17\approx 0.941$ of Austrin, O'Donnell, Tan and Wright, the best NP-hardness
result for \MC{3} since 2012. (Their result, unlike ours, also holds for $3$-colourable graphs.) We also
improve the NP-hardness of \MC{k} for every $k$ from $4$ to $20$; for instance, we obtain $45/47\approx0.957$
for \MC{4} and $33/34\approx0.971$ for \MC{5}, where the previous factors were $0.987$ and $120/121$.

Our reductions are gadget reductions from H\aa stad's linear equations over $\Z_3$, with gadgets that have
at most six auxiliary vertices. The main new idea is a \emph{negation trick}: since the
negation of an assignment is again an assignment, equations with right-hand sides $t$ and $-t$ can be
handled together, and this makes the resulting predicates symmetric under all permutations of the
colours. We complement it with a general soundness criterion for such reductions: in effect, the best
strategy of the adversary is to use a single colour pattern everywhere. This criterion is the strongest
analysis that uses H\aa stad's theorem as a black box, and, using LP duality, we show that under it no
gadget reduction from these equations can beat $27/29$ for \MC{3}.

Finally, we observe that a classical reduction from \MC{3} to \MC{k}, applied more carefully, already
gives $82/85\approx0.965$ for \MC{4} on $4$-colourable graphs, improving the previously stated $85/86$.
\end{abstract}

\newpage
\setcounter{tocdepth}{2}
\tableofcontents
\thispagestyle{empty}
\newpage
\setcounter{page}{1}

\section{Introduction}

In \MC{k}, we are given a graph with non-negative edge weights and must colour its vertices with $k$
colours so as to maximise the total weight of edges whose endpoints receive different colours. For $k=2$
this is \textsc{Max-Cut}. For $k\ge 3$ the best approximation algorithms round the semidefinite
relaxation of Frieze and Jerrum~\cite{FJ97}; they achieve a ratio of about $0.836$ for $k=3$ and $0.857$
for $k=4$~\cite{GW04,dKPW04}. For $k=3$ this is optimal under the Unique Games
Conjecture~\cite{IM12,Heilman26}.

Much less is known about NP-hardness. Kann, Khanna, Lagergren and Panconesi~\cite{KKLP97} proved hardness
of the form $1-\Theta(1/k)$, and Guruswami and Sinop~\cite{GS13} proved bounds of the same form even for
$k$-colourable graphs. The current
record for $k=3$ was set by Austrin, O'Donnell, Tan and Wright~\cite{AOTW14}: using a PCP designed for the
purpose, they showed that \MC{3} is NP-hard to approximate within $16/17+\eps$, even on $3$-colourable
graphs. Combined with a reduction of Guruswami and Sinop, the same result gives bounds for every
$k\ge4$, such as $85/86$ for \MC{4}~\cite{AOTW14}; on $k$-colourable graphs, and on general graphs for
$k\ge5$, these are the best previously stated bounds. Most recently, Nagda, Raghavan
and Thakurta~\cite{NRT25} used the AlphaEvolve system~\cite{AlphaEvolve} to search for gadget reductions
from H\aa stad's linear equations~\cite{Hastad01}. This improved the bound for \MC{4} to about $0.987$,
but for \MC{3} their gadgets give only $55/57\approx0.965$, short of $16/17$. \Cref{tab:summary} collects
these bounds.

\subsection{Our results}

Our first result improves the $16/17$ bound for \MC{3}.

\begin{theorem}\label{thm:main3}
For every $\eps>0$ it is NP-hard, given a weighted graph $G$, to distinguish between the case that some
$3$-colouring cuts at least a $(29/30-\eps)$ fraction of the total edge weight and the case that every
$3$-colouring cuts at most a $(9/10+\eps)$ fraction. In particular, \MC{3} is NP-hard to approximate
within $27/29+\eps\approx 0.931$.
\end{theorem}

The same approach, with a few extra vertices added to the gadgets, gives improved bounds for every $k$.

\begin{theorem}\label{thm:main45}
For every $k\ge3$ and $\eps>0$, \MC{k} is NP-hard to approximate within $1-2/\Gamma_{k-3}+\eps$, where
$\Gamma_L=29+18L+3\binom L2$. In particular, \MC{4} is NP-hard to approximate within $45/47+\eps\approx
0.957$, and \MC{5} within $33/34+\eps\approx0.971$.
\end{theorem}

For $k=3$, \cref{thm:main45} is \cref{thm:main3}. Unlike the result of~\cite{AOTW14}, our reductions
do not have perfect completeness: on $3$-colourable graphs, $16/17$ remains the best known bound. For
$k$-colourable graphs with $k\ge4$, however, we observe that the bounds previously stated in the
literature can also be improved, by applying a known reduction more carefully.

\begin{theorem}\label{thm:perfect}
For every $\eps>0$ it is NP-hard to distinguish $4$-colourable graphs from graphs in which every
$4$-colouring cuts at most an $82/85+\eps\approx0.965$ fraction of the edge weight, and $5$-colourable
graphs from graphs in which every $5$-colouring cuts at most a $119/122+\eps\approx0.975$ fraction.
Similar improvements hold for every $k$ from $4$ to $20$ (\cref{sec:universal}).
\end{theorem}

\begin{figure}[t]
\centering
\begingroup
\begin{tikzpicture}
\begin{axis}[
  width=0.8\textwidth, height=5.5cm,
  scale only axis=false,
  xmin=2.5, xmax=24.5,
  ymode=log, ymin=0.0016, ymax=0.09,
  xtick={3,...,24},
  ytick={0.002,0.005,0.01,0.02,0.05},
  yticklabels={$0.002$,$0.005$,$0.01$,$0.02$,$0.05$},
  minor ytick={0.003,0.004,0.006,0.007,0.008,0.009,0.03,0.04,0.06,0.07,0.08,0.09},
  ymajorgrids, yminorgrids=false,
  major grid style={black!14,line width=0.3pt},
  tick label style={font=\footnotesize},
  tick style={black!70},
  xlabel={$k$}, xlabel style={font=\small,yshift=3pt},
  ylabel={$1-{}$hardness factor\\[-2pt]{\footnotesize(larger is stronger)}},
  ylabel style={font=\small,align=center,yshift=-2pt},
  axis on top,
  legend cell align=left,
  legend pos=north east,
  legend style={font=\footnotesize,draw=black!40,fill=white,inner sep=2pt,
                row sep=-1pt,line width=0.4pt},
  every axis plot/.append style={line join=round},
]
\fill[black!6] (axis cs:2.5,0.0016) rectangle (axis cs:20.5,0.09);
\node[font=\footnotesize,text=black!60,anchor=south west,inner sep=0pt]
  at (axis cs:4.7,0.00218) {shaded: $k\le 20$, improved by this paper};
\addplot[blue!60!black, line width=0.8pt, mark=*, mark size=1.5pt,
         mark options={solid,fill=blue!60!black}]
  coordinates {(3,0.0689655) (4,0.0425532) (5,0.0294118) (6,0.0217391) (7,0.0168067)
   (8,0.0134228) (9,0.010989) (10,0.00917431) (11,0.0077821) (12,0.00668896)
   (13,0.00581395) (14,0.00510204) (15,0.00451467) (16,0.00402414) (17,0.00361011)
   (18,0.00325733) (19,0.00295421) (20,0.00269179) (21,0.00246305) (22,0.00226244)
   (23,0.00208551) (24,0.00192864)};
\addlegendentry{this paper (Thm~\ref{thm:main45})}
\addplot[black, line width=0.6pt, dashed, mark=square*, mark size=1.3pt,
         mark options={solid,fill=white}]
  coordinates {(4,0.0352941) (5,0.0245902) (6,0.0185185) (7,0.0146341) (8,0.0119522)
   (9,0.01) (10,0.00852273) (11,0.00737101) (12,0.00645161) (13,0.00570342)
   (14,0.00508475) (15,0.00456621) (16,0.00412655) (17,0.00375) (18,0.00342466)
   (19,0.00314136) (20,0.00289296) (21,0.0026738) (22,0.00247934) (23,0.00230592)
   (24,0.00215054)};
\addlegendentry{this paper, $k$-colourable (Thm~\ref{thm:perfect})}
\addplot[black!65, line width=0.6pt, densely dotted, mark=triangle*, mark size=1.7pt,
         mark options={solid,fill=black!65}]
  coordinates {(3,0.0588235) (4,0.0116279) (5,0.00826446) (6,0.00980392) (7,0.00729927)
   (8,0.00581395) (9,0.00653595) (10,0.00531915) (11,0.0044843) (12,0.00490196)
   (13,0.0041841) (14,0.00364964) (15,0.00392157) (16,0.00344828) (17,0.00307692)
   (18,0.00326797) (19,0.00293255) (20,0.00265957) (21,0.00280112) (22,0.00255102)
   (23,0.00234192) (24,0.00245098)};
\addlegendentry{previous \cite{AOTW14}}
\addplot[only marks, black, mark=diamond*, mark size=2.1pt, mark options={fill=white}]
  coordinates {(4,0.01292)};
\addlegendentry{\cite{NRT25}}
\end{axis}
\end{tikzpicture}%
\endgroup
\caption{NP-hardness of approximation for \MC{k}, $3\le k\le24$, plotted as $1-{}$(hardness factor) on a
logarithmic scale, so higher points are stronger bounds. The plotted values are $2/\Gamma_{k-3}$ for
\cref{thm:main45} (general graphs), $3/(51+34L+3\binom L2)$ with $L=k-3$ for \cref{thm:perfect}
($k$-colourable graphs, $k\ge4$), and $1/(17(k+c_k)+c_k)$ with $c_k=k\bmod3$ for the previous bound
of~\cite{AOTW14} ($1/17$ for $k=3$); the diamond is the bound $0.987$ of~\cite{NRT25} for \MC{4}.
\Cref{thm:main45} is the strongest of the three for $k\le14$, \cref{thm:perfect} for $15\le k\le20$,
and~\cite{AOTW14} for $k\ge21$; in the shaded range $k\le20$ this paper improves the best previous bound.}
\label{fig:bounds}
\end{figure}
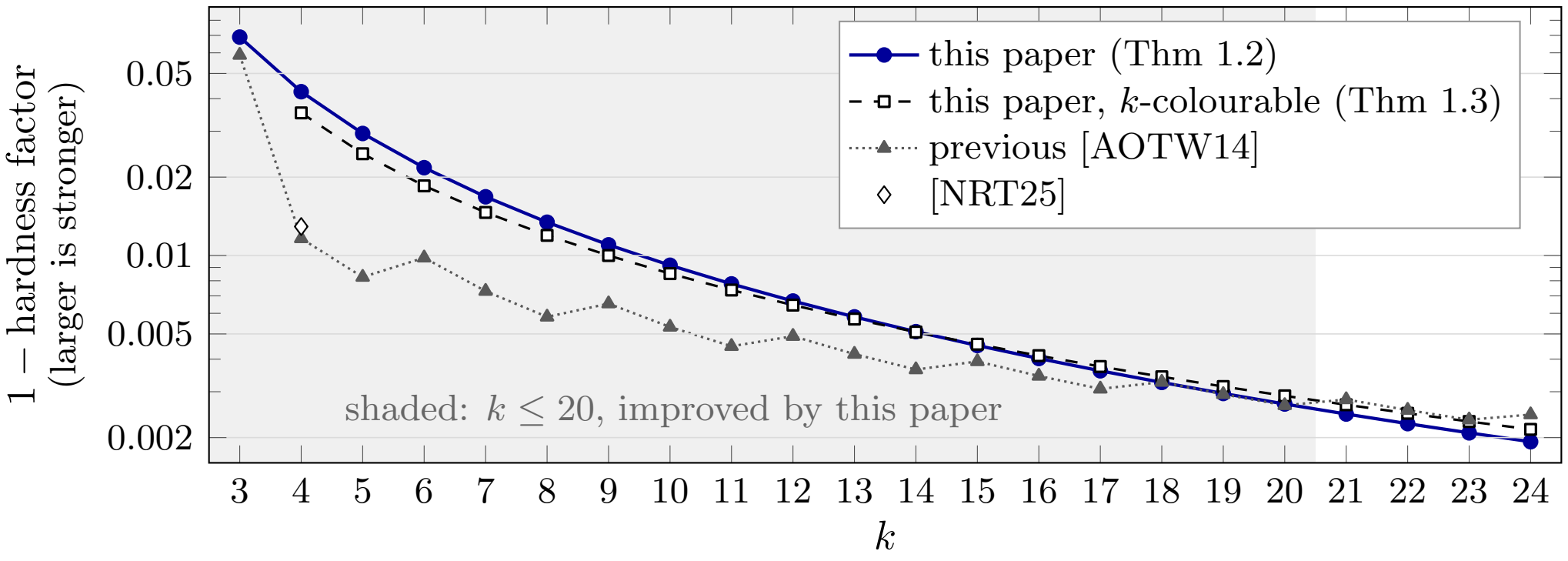

\Cref{tab:summary} and \cref{fig:bounds} compare our bounds with the previous ones. Together, \cref{thm:main45,thm:perfect}
improve the best known NP-hardness of \MC{k} for every $k$ from $3$ to $20$; for larger $k$ the bound
of~\cite{AOTW14} remains the best. All three theorems have short proofs by hand.

\begin{table}[t]
\centering\small\setlength{\tabcolsep}{5pt}
\begin{tabular}{@{}lcccc@{}}
\toprule
& \multicolumn{2}{c}{on $k$-colourable graphs} & \multicolumn{2}{c}{on general graphs} \\
\cmidrule(lr){2-3}\cmidrule(l){4-5}
& previous~\cite{AOTW14} & \cref{thm:perfect} & previous & \cref{thm:main3,thm:main45} \\
\midrule
$k=3$ & $16/17\approx 0.941$ & --- & $16/17$~\cite{AOTW14} & $27/29 \approx 0.931$ \\
$k=4$ & $85/86\approx 0.988$ & $82/85\approx 0.965$ & $0.987$~\cite{NRT25} & $45/47\approx 0.957$ \\
$k=5$ & $120/121\approx 0.992$ & $119/122\approx 0.975$ & $120/121$~\cite{AOTW14} & $33/34\approx 0.971$ \\
$k=6$ & $101/102\approx 0.990$ & $53/54\approx 0.981$ & $101/102$~\cite{AOTW14} & $45/46 \approx 0.978$ \\
$k=7$ & $136/137\approx 0.993$ & $202/205\approx 0.985$ & $136/137$~\cite{AOTW14} & $117/119\approx 0.983$ \\
\bottomrule
\end{tabular}
\caption{NP-hardness of approximation for \MC{k}; smaller factors are stronger. The previous values for
$k\ge4$ are instances of the general formula $1-\frac1{17(k+c_k)+c_k}$, $c_k=k\bmod3$,
of~\cite[Thm.~1.2]{AOTW14}, except the value $0.987$ of~\cite{NRT25}.}
\label{tab:summary}
\end{table}

\subsection{Overview of the techniques}

\paragraph{Gadget reductions and the colour-reference problem.}
H\aa stad~\cite{Hastad01} proved that for systems of equations $x_1+x_2+x_3 = t$ over $\Z_3$ it is NP-hard
to distinguish almost satisfiable systems from systems in which every assignment satisfies at most a
$1/3+\eps$ fraction of the equations. The classical way to turn this into hardness for a graph problem,
introduced by Trevisan, Sorkin, Sudan and Williamson~\cite{TSSW00} for $\Z_2$ and used for $\Z_3$
by~\cite{NRT25}, replaces each equation by a small graph, a \emph{gadget} (\cref{fig:reduction}), whose best cut is larger when
the equation is satisfied. For $t\ne0$, however, the predicate ``$x_1+x_2+x_3=t$'' is not invariant under
renaming the colours, so its gadget must refer to a fixed set of \emph{global} vertices that play the role
of the colours $0,1,2$. In a cheating colouring all global vertices may receive the same colour, and
guarding against this makes the gadgets for $t\ne0$ large and inefficient; in the analysis
of~\cite{NRT25} they are the bottleneck that keeps the bound above $16/17$.

\paragraph{The negation trick.}
Our main observation is that the negation $-X$ of an assignment $X$ is again an assignment, and that $-X$
satisfies an equation with right-hand side $t$ exactly when $X$ satisfies the same equation with
right-hand side $-t$. If every assignment satisfies at most a $1/3+\eps$ fraction of the equations, then
applying this to both $X$ and $-X$ shows that, for a random equation, the sum $s$ of its variables under
$X$ satisfies $s=t$ or $s=-t$ with probability at most about $2/3$. It is therefore enough for the gadgets of
the equations with $t=\pm1$ to detect whether $s\ne0$, which simply means that exactly two of the three variables are equal (\cref{fig:tuples}). This predicate is invariant under all permutations of the colours, needs no global
vertices, and is detected by a triangle together with one extra vertex (the gadget $N$ of \cref{fig:gadgets}(b)). Combined with the gadget
of~\cite{NRT25} for $t=0$, this proves \cref{thm:main3} in a few lines (\cref{sec:three}).

\paragraph{A soundness criterion.}
For $k\ge4$ the satisfying colourings use only three of the $k$ colours, and global vertices become
necessary again. We prove a general criterion for gadget reductions from H\aa stad's equations
(\cref{thm:S}): under a mild symmetry condition, the gadgets give NP-hardness of approximation within
$\rho+\delta$ for every $\delta>0$, where
\[
  \rho \;=\; \frac{\max_{x,y}\; \sum_{t} V_t(x,y)}{\sum_{t} C_t},
\]
$V_t(x,y)$ is the value of the gadget for right-hand side $t$ when its three variables receive the
colours $x$ and the global vertices the colours $y$, and $C_t$ is its least value on satisfying colourings. In
words, a cheating colouring can do no better than to use a single colour pattern $x$ on all equations,
and each pattern is charged with the sum of the values of all three gadgets. The proof decodes a
colouring of the graph into many assignments of H\aa stad's instance and uses a short Fourier-analytic
argument; it relies on the fact that H\aa stad's equations have one variable of one kind and two of
another. We also show that no analysis that uses H\aa stad's theorem only as a black box can do better
(\cref{prop:tight}). The criterion contains the negation trick and, for $k=3$, the analysis
of~\cite{NRT25} as special cases. It also turns gadget design into a single linear program over the three
gadgets. Solving this program for $k=4,5$ and simplifying the solutions by hand led to the gadgets behind
\cref{thm:main45}: the gadgets of \cref{thm:main3} together with $k-3$ global vertices of the extra colours
(\cref{sec:family}).

\paragraph{Universal vertices.}
\Cref{thm:perfect} uses a reduction of~\cite{KKLP97,GS13} that adds vertices joined to the whole graph
and forces them to take the extra colours. The previously stated constants apply it on top of a general
bound for $K$-colourable graphs ($K$ divisible by $3$) which, for $K=3$, is three times weaker than the
$16/17$ bound it is derived from; applying it directly to the $16/17$ bound gives \cref{thm:perfect}
(\cref{sec:universal}). Placing the extra vertices
inside the gadgets, as in \cref{thm:main45}, is more economical than adding them to the final graph.

\paragraph{A barrier.}
Finally, we show that $27/29$ is the best factor that the criterion can give for \MC{3}, for gadgets of
any size (\cref{thm:barrier}): there is a single distribution of colour patterns on which every gadget,
for every right-hand side, still achieves at least a $27/29$ fraction of its completeness value on
average (\cref{fig:barrier}). The proof uses LP
duality, in the spirit of the optimality proofs of~\cite{TSSW00,HHMOW17}, with explicit rational
certificates that we verify by computer in exact arithmetic; this is the only computer-assisted result
of the paper. To go beyond $27/29$, one has to change the source problem (for instance, to other
predicates covered by Chan's theorem~\cite{Chan16}), the form of the reduction, or the analysis.

\paragraph{How these results were obtained.}
This research was carried out with extensive use of an AI system, which designed and ran the computer
searches, found the arguments and drafted the paper under the direction of the author. The methodology,
including how every claim was checked, is described in \cref{app:comp} and in the AI methodology
statement at the end of the paper.

\paragraph{Organisation.}
\Cref{sec:prelim} fixes notation and states H\aa stad's theorem in the form we use. \Cref{sec:three} proves
\cref{thm:main3}, \cref{sec:S} the soundness criterion, and \cref{sec:family} \cref{thm:main45}.
\Cref{sec:universal} treats universal vertices and \cref{thm:perfect}, \cref{sec:barrier} the barrier, and
\cref{sec:discussion} open problems. \Cref{app:remarks} collects further remarks on the soundness criterion,
and \cref{app:comp} describes the computations.

\section{Preliminaries}\label{sec:prelim}

\paragraph{Cuts.} We write $[k]=\{0,1,\dots,k-1\}$ and identify $[3]$ with $\Z_3$. A (weighted) graph
$G=(V,E,w)$ has non-negative edge weights; for a colouring $\chi\colon V\to[k]$ let
$\cut_\chi(G)=\sum_{\{u,v\}\in E} w_{uv}\,[\chi(u)\neq\chi(v)]$ and $w(G)=\sum_{\{u,v\}\in E} w_{uv}$. We say that
\MC{k} is $(c,s)$-hard if it is NP-hard to distinguish $\max_\chi \cut_\chi(G)\ge c\,w(G)$ from
$\max_\chi \cut_\chi(G)\le s\, w(G)$; this implies NP-hardness of approximation within $s/c$. Weights may
be removed at an arbitrarily small loss by the method of~\cite{CST01}; see~\cite[Remark~2.6]{GS13} for
\MC{k}. The transformation replaces vertices by independent sets and edges by bipartite graphs, so it
also preserves $k$-colourability; hence all results hold for unweighted simple graphs.

\paragraph{H\aa stad's equations.}
A \emph{role-separated} instance $\Phi$ of $\lin$ consists of disjoint sets $U$ and $W$ of variables and a
collection of equations $e$ of the form $u_e+w_e+w'_e = t_e$ over $\Z_3$, with $u_e\in U$, distinct
$w_e,w'_e\in W$, $t_e\in\Z_3$ (the \emph{type} of $e$), and weights $\lambda_e\ge0$ summing to $1$. For an
assignment $\sigma\colon U\cup W\to\Z_3$, $\val_\Phi(\sigma)$ is the total weight of the equations it
satisfies, and $\val(\Phi)=\max_\sigma\val_\Phi(\sigma)$. $\Phi$ is \emph{balanced} if each type has total
weight $1/3$.

\begin{theorem}[H\aa stad~\cite{Hastad01}]\label{thm:hastad}
For every $\eps>0$ it is NP-hard, given a balanced role-separated instance $\Phi$ of $\lin$, to
distinguish $\val(\Phi)\ge 1-\eps$ from $\val(\Phi)\le 1/3+\eps$.
\end{theorem}

\begin{proof}[Comments on the statement]
This is H\aa stad's theorem for the group $\Z_3$~\cite[Thm.~5.9]{Hastad01}; the gap version follows from
Lemmas~5.10 and~5.11 there. In his verifier, each test reads one entry of a folded long-code table indexed
by a set $U$ of variables and two entries of a table indexed by a set $W$ of clauses, and accepts iff the
three values sum to a constant; folding only changes the right-hand side. Giving the two provers separate
tables makes the instance role-separated; the soundness analysis~\cite[Lemma~5.11]{Hastad01} applies to
arbitrary tables, i.e.\ to arbitrary assignments, and constructs the strategies of the two provers from
the two kinds of tables separately. Tests whose two $W$-entries fall into the same folded class have
negligible total weight (it tends to $0$ with the parameter of the verifier); we discard them, which
changes completeness and soundness by a negligible amount.

To balance the types, replace every $u\in U$ by three copies $u^{(0)},u^{(1)},u^{(2)}$ and each equation $e$
by the three equations $u_e^{(c)}+w_e+w'_e = t_e+c$ ($c\in\Z_3$), each of weight $\lambda_e/3$. The new
instance $\Phi'$ is role-separated (with $U\times\Z_3$ and $W$), and every type has weight exactly $1/3$. An
assignment $\sigma$ of $\Phi$ yields $\sigma(u^{(c)})=\sigma(u)+c$, which satisfies the same weight.
Conversely, for any assignment $\tau$ of $\Phi'$,
\[
\val_{\Phi'}(\tau)=\tfrac13\textstyle\sum_{c\in\Z_3}\val_\Phi(\sigma_c),\qquad
\sigma_c(u)=\tau(u^{(c)})-c,\ \ \sigma_c(w)=\tau(w),
\]
and each $\sigma_c$ is an assignment of $\Phi$ (the values on $W$ are shared), so $\val(\Phi')\le\val(\Phi)$.
\end{proof}

\paragraph{Gadgets.}
Fix $k\ge 3$. A \emph{gadget} $I$ is a weighted graph on three \emph{primary} vertices $p_1,p_2,p_3$,
$k$ \emph{global} vertices $g_0,\dots,g_{k-1}$ and some \emph{auxiliary} vertices. For $x\in[k]^3$ and
$y\in[k]^k$ let
\[
  V_I(x,y) = \max\{\cut_\chi(I) : \chi(p_j)=x_j,\ \chi(g_i)=y_i\},
\]
the maximum being over the colours of the auxiliary vertices. Let $\yc=(0,1,\dots,k-1)$ be the canonical
colouring of the globals. The value of a colouring does not change when the $k$ colours are permuted, so
we may normalise the colours of the globals: let $\Y_k\subseteq[k]^k$ contain one representative of each
orbit under permutations of the colours (we use restricted growth strings: the first entry is $0$, and
each entry exceeds the maximum of the previous ones by at most $1$). Since $V_I(\pi x,\pi y)=V_I(x,y)$ for every
permutation $\pi$ of $[k]$, maxima of $V_I$, or of sums of several $V_I$, jointly over $x\in[k]^3$ and
$y\in\Y_k$ equal the maxima over $x\in[k]^3$ and all $y\in[k]^k$; the normalisation is only a
computational saving. A \emph{family} for $\lin$ is a triple $(I_0,I_1,I_2)$ of gadgets, $I_t$ to be used for equations of
type $t$. We write $V_t=V_{I_t}$ and $S_y(x)=\sum_tV_t(x,y)$. Let
$\sat_t=\{x\in\Z_3^3: x_1+x_2+x_3=t\}$ be the set of \emph{satisfying} tuples of type $t$; they use only the
colours $0,1,2$. The \emph{completeness} of $I_t$ is
\[
  C_t = \min\{V_t(x,\yc) : x\in\sat_t\}.
\]

\begin{figure}[t]
\centering
\begin{tikzpicture}[x=1cm,y=1cm,
  prim/.style={circle,draw,fill=black!15,inner sep=1pt,minimum size=14pt,font=\small},
  aux/.style={circle,draw,fill=white,inner sep=0pt,minimum size=4.6pt,line width=0.4pt},
  univ/.style={rectangle,draw,fill=black!5,inner sep=0pt,minimum size=9pt},
  gad/.style={draw=black!60,fill=black!4,rounded corners=4pt,line width=0.5pt,
              minimum width=4.4cm,minimum height=0.68cm,inner sep=0pt},
  leg/.style={line width=0.5pt,black!60},
  glob/.style={dash pattern=on 2.6pt off 1.6pt,blue!60!black,line width=0.7pt},
  tap/.style={circle,fill=blue!60!black,inner sep=0pt,minimum size=3.2pt},
  gname/.style={font=\small,inner sep=1pt},
  eqn/.style={font=\footnotesize,inner sep=1pt,anchor=west},
  plab/.style={font=\scriptsize,text=black!55,inner sep=1.5pt},
  clab/.style={font=\small,text=black!60,inner sep=1pt},
  dlab/.style={font=\footnotesize,text=black!50,inner sep=1pt}]
\def\xu{0}\def\xw{10.3}\def\bx{5.15}
\foreach \n/\x/\l in {0/3.45/g_0, 1/4.25/g_1, 2/5.05/g_2, 3/6.85/g_{k-1}}{
  \node[univ] (g\n) at (\x,4.85) {};
  \node[font=\footnotesize,inner sep=1pt,anchor=south] at ([yshift=1pt]g\n.north) {$\l$};}
\node[font=\small] at (5.95,4.85) {$\cdots$};
\draw[decorate,decoration={brace,mirror,amplitude=4pt},black!45,line width=0.5pt]
  (3.25,4.6) -- (7.05,4.6);
\node[dlab,anchor=west] at (7.3,4.85) {shared by all copies};
\draw[glob] (\bx,4.44) -- (\bx,1.6);
\node[prim] (u1) at (\xu,3.1) {$u_1$};
\node[prim] (u2) at (\xu,1.6) {$u_2$};
\node[prim] (w1) at (\xw,4.1) {$w_1$};
\node[prim] (w2) at (\xw,3.1) {$w_2$};
\node[prim] (w3) at (\xw,2.1) {$w_3$};
\node[prim] (w4) at (\xw,1.1) {$w_4$};
\foreach \n/\y/\t/\eq in {0/3.6/0/{u_1+w_1+w_2=0}, 1/2.6/1/{u_1+w_2+w_3=1}, 2/1.6/2/{u_2+w_3+w_4=2}}{
  \node[gad] (B\n) at (\bx,\y) {};
  \node[gname] at ([xshift=0.36cm]B\n.west) {$I_{\t}$};
  \draw[black!30,line width=0.4pt]
    ([xshift=0.68cm,yshift=0.19cm]B\n.west) -- ([xshift=0.68cm,yshift=-0.19cm]B\n.west);
  \node[eqn] at ([xshift=0.78cm]B\n.west) {$\eq$};
  \node[aux] (x\n a) at ([xshift=-0.80cm,yshift=0.11cm]B\n.east) {};
  \node[aux] (x\n b) at ([xshift=-0.52cm,yshift=-0.11cm]B\n.east) {};
  \node[aux] (x\n c) at ([xshift=-0.24cm,yshift=0.11cm]B\n.east) {};
  \draw[black!45,line width=0.3pt] (x\n a) -- (x\n b) -- (x\n c) -- (x\n a);
  \node[tap] at (B\n.north) {};}
\draw[leg] (B0.west) -- node[plab,pos=0.09,above] {$p_1$} (u1);
\draw[leg] (B1.west) -- (u1);
\draw[leg] (B2.west) -- (u2);
\draw[leg] ([yshift=0.16cm]B0.east) -- node[plab,pos=0.09,above] {$p_2$} (w1);
\draw[leg] ([yshift=-0.16cm]B0.east) -- node[plab,pos=0.09,below] {$p_3$} (w2);
\draw[leg] ([yshift=0.16cm]B1.east) -- (w2);
\draw[leg] ([yshift=-0.16cm]B1.east) -- (w3);
\draw[leg] ([yshift=0.16cm]B2.east) -- (w3);
\draw[leg] ([yshift=-0.16cm]B2.east) -- (w4);
\node[clab] at (\xu,0.6) {$U$};
\node[dlab] at (\xu,0.27) {decoded by $a$};
\node[clab] at (\xw,0.6) {$W$};
\node[dlab] at (\xw,0.27) {decoded by $b$};
\end{tikzpicture}
\caption{The graph $J(\Phi)$ for a balanced role-separated instance $\Phi$ with three equations of
weight $\frac13$, one of each type. Each variable is a single vertex, shared by the equations that contain
it ($u_1$, $w_2$ and $w_3$ occur twice). Each equation $e$ contributes a copy of $I_{t_e}$ scaled by
$\lambda_e$ (box) with new auxiliary vertices (small circles, schematic); its primaries $p_1,p_2,p_3$
(labelled on the top copy) are identified with $u_e,w_e,w'_e$ (grey lines, which mark identifications and are not edges of $J$) and its globals with the shared vertices
$g_0,\dots,g_{k-1}$ (dashed line, drawn behind the boxes), so different copies meet only in variable
vertices and globals. \Cref{sec:S} decodes a colouring $\chi$ into the assignment $u\mapsto a(\chi(u))$ on
$U$, $w\mapsto b(\chi(w))$ on $W$, for arbitrary maps $a,b\colon[k]\to\Z_3$.}
\label{fig:reduction}
\end{figure}
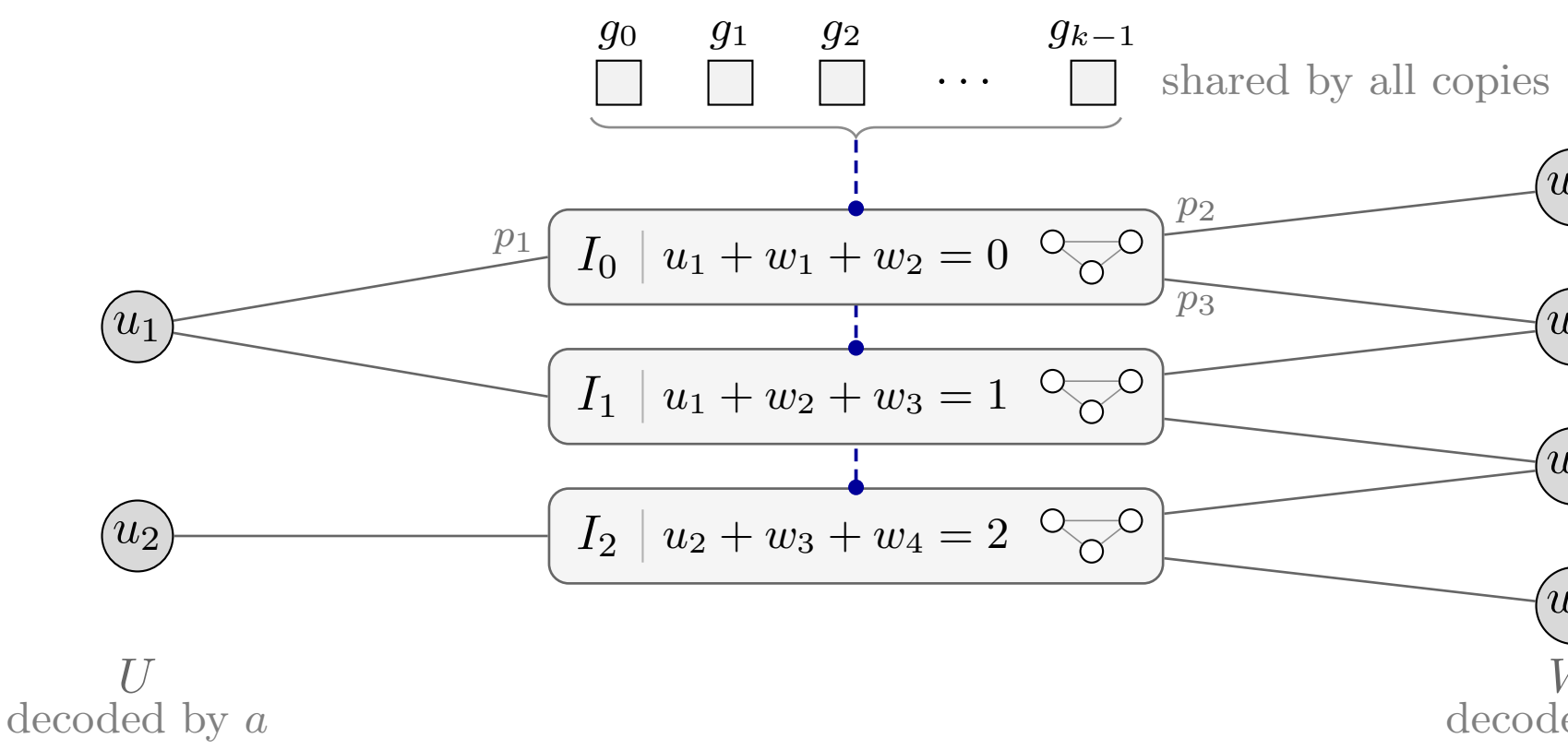

\paragraph{The reduction.}
Given a role-separated $\Phi$ and a family, the graph $J=J(\Phi)$ (\cref{fig:reduction}) has a vertex for every variable of
$\Phi$, $k$ global vertices, and for every equation $e$ a copy of $I_{t_e}$ scaled by $\lambda_e$, whose
primaries $p_1,p_2,p_3$ are identified with $u_e,w_e,w'_e$, whose global vertices are identified with the
global vertices of $J$, and whose auxiliary vertices are new. For balanced $\Phi$,
$w(J)=\frac13\sum_t w(I_t)$. If $\val(\Phi)\ge 1-\eps$ and $\Phi$ is balanced, colouring the variables by
an assignment that satisfies weight at least $1-\eps$, the globals by $\yc$, and the auxiliary vertices of
each satisfied equation optimally gives
\begin{equation}\label{eq:completeness}
  \max_\chi \cut_\chi(J) \;\ge\; \tfrac13\textstyle\sum_t C_t \;-\; \eps\max_t C_t .
\end{equation}
For every colouring $\chi$ of $J$, after normalising $y=\chi(g_0,\dots,g_{k-1})\in\Y_k$,
\begin{equation}\label{eq:soundness-basic}
  \cut_\chi(J) \;\le\; \sum_e \lambda_e\, V_{t_e}\big(\chi(u_e),\chi(w_e),\chi(w'_e),\,y\big),
\end{equation}
since the auxiliary vertices of different copies are disjoint.

\section{Max-3-Cut: the negation trick}\label{sec:three}

\begin{figure}[t]
\centering
\begin{tikzpicture}[x=1cm,y=1cm,
  cls/.style={draw,rounded corners=2pt,line width=0.5pt},
  tup/.style={font=\small,inner sep=0pt},
  hl/.style={tup,text=blue!60!black},
  cname/.style={font=\small,inner sep=1pt,text height=1.6ex,text depth=0.5ex},
  cdesc/.style={font=\footnotesize,text=black!70,inner sep=1pt,text height=1.6ex,text depth=0.5ex},
  rlab/.style={font=\small,anchor=east,inner sep=1pt},
  cell/.style={draw=black!45,line width=0.4pt},
  hi/.style={cell,fill=black!12},
  val/.style={font=\small},
  neg/.style={<->,>={Stealth[length=4pt,width=3.5pt]},blue!60!black,line width=0.6pt},
  brc/.style={decorate,decoration={brace,amplitude=4pt},line width=0.5pt},
  note/.style={font=\footnotesize,anchor=west,text=black!70,inner sep=1pt},
  slab/.style={above=4pt,font=\small,inner sep=1pt,text height=1.7ex,text depth=0.5ex}]
\def\rh{0.46}\def\bt{1.58}
\foreach \l/\r in {0/1, 1.75/3.75, 4.5/7.5, 8.25/11.25}
  \draw[cls] (\l,0) rectangle (\r,\bt);
\foreach \i/\a/\d/\e/\p/\q/\r/\pp/\qq/\rr in {
    0/000/012/021/001/010/100/002/020/200,
    1/111/102/201/112/{}/211/221/{}/122,
    2/222/120/210/022/202/220/011/101/110}{
  \path (0,\bt-0.1-\i*\rh-0.5*\rh) coordinate (r);
  \node[tup] at ([xshift=0.5cm]r)   {$\a$};
  \node[tup] at ([xshift=2.25cm]r)  {$\d$};
  \node[tup] at ([xshift=3.25cm]r)  {$\e$};
  \node[tup] at ([xshift=5cm]r)     {$\p$};
  \node[tup] at ([xshift=6cm]r)     {$\q$};
  \node[tup] at ([xshift=7cm]r)     {$\r$};
  \node[tup] at ([xshift=8.75cm]r)  {$\pp$};
  \node[tup] at ([xshift=9.75cm]r)  {$\qq$};
  \node[tup] at ([xshift=10.75cm]r) {$\rr$};}
\node[hl] at (6,\bt-0.1-1.5*\rh)    {$\mathbf{121}$};
\node[hl] at (9.75,\bt-0.1-1.5*\rh) {$\mathbf{212}$};
\foreach \c/\n/\d in {0.5/{\mathcal A}/all equal, 2.75/{\mathcal D}/all distinct,
                      6/{\mathcal T_1}/{two equal, $s=1$}, 9.75/{\mathcal T_2}/{two equal, $s=2$}}{
  \node[cname] at (\c,\bt+0.62) {$\n$};
  \node[cdesc] at (\c,\bt+0.25) {\d};}
\draw[brc] (0,\bt+0.92)    -- node[slab] {$\sat_0$\ \ ($s(x)=0$)} (3.75,\bt+0.92);
\draw[brc] (4.5,\bt+0.92)  -- node[slab] {$\sat_1$} (7.5,\bt+0.92);
\draw[brc] (8.25,\bt+0.92) -- node[slab] {$\sat_2$} (11.25,\bt+0.92);
\node[rlab] at (-0.2,-0.3) {$x\mapsto -x$};
\draw[neg] (0.3,-0.06)  to[out=-60,in=-120,looseness=1.6] (0.7,-0.06);
\draw[neg] (2.45,-0.06) to[out=-50,in=-130,looseness=1.3] (3.05,-0.06);
\draw[neg] (6,-0.06)    to[out=-14,in=-166] (9.75,-0.06);
\def\yA{-0.62}\def\yB{-1.12}\def\yC{-1.18}\def\yD{-1.68}
\node[rlab] at (-0.2,{(\yA+\yB)/2}) {$V_E$};
\node[rlab] at (-0.2,{(\yC+\yD)/2}) {$V_N$};
\foreach \l/\r/\v/\s in {0/1/18/hi, 1.75/3.75/18/hi, 4.5/7.5/16/cell, 8.25/11.25/16/cell}
  \draw[\s] (\l,\yA) rectangle node[val] {$\v$} (\r,\yB);
\foreach \l/\r/\v/\s in {0/1/9/cell, 1.75/3.75/9/cell, 4.5/7.5/11/hi, 8.25/11.25/11/hi}
  \draw[\s] (\l,\yC) rectangle node[val] {$\v$} (\r,\yD);
\node[note] at (11.45,{(\yA+\yB)/2}) {larger iff $s(x)=0$};
\node[note] at (11.45,{(\yC+\yD)/2}) {larger iff $s(x)\neq0$};
\end{tikzpicture}
\caption{The $27$ tuples $x=x_1x_2x_3\in\Z_3^3$ in the classes $\mathcal A$ (all equal), $\mathcal D$
(all distinct), $\mathcal T_1$, $\mathcal T_2$ (two equal, $s(x)=x_1+x_2+x_3=1$ resp.\ $2$); the tuples with $s(x)=t$ form $\sat_t$, so
$\sat_0=\mathcal A\cup\mathcal D$, $\sat_1=\mathcal T_1$ and $\sat_2=\mathcal T_2$. Negation $x\mapsto-x$
maps $\mathcal A$ and $\mathcal D$ to themselves (on $\mathcal D$ it exchanges the two columns) and maps
each tuple of $\mathcal T_1$ to the tuple of $\mathcal T_2$ at the same position (e.g.\ $121\mapsto212$).
The bottom rows give the gadget values of \cref{lem:EN}, with the larger value of each gadget shaded:
$V_E$ is larger exactly on $\sat_0$, and $V_N$ exactly on $\sat_1\cup\sat_2$, so $N$ cannot tell
$\mathcal T_1$ from $\mathcal T_2$.}
\label{fig:tuples}
\end{figure}
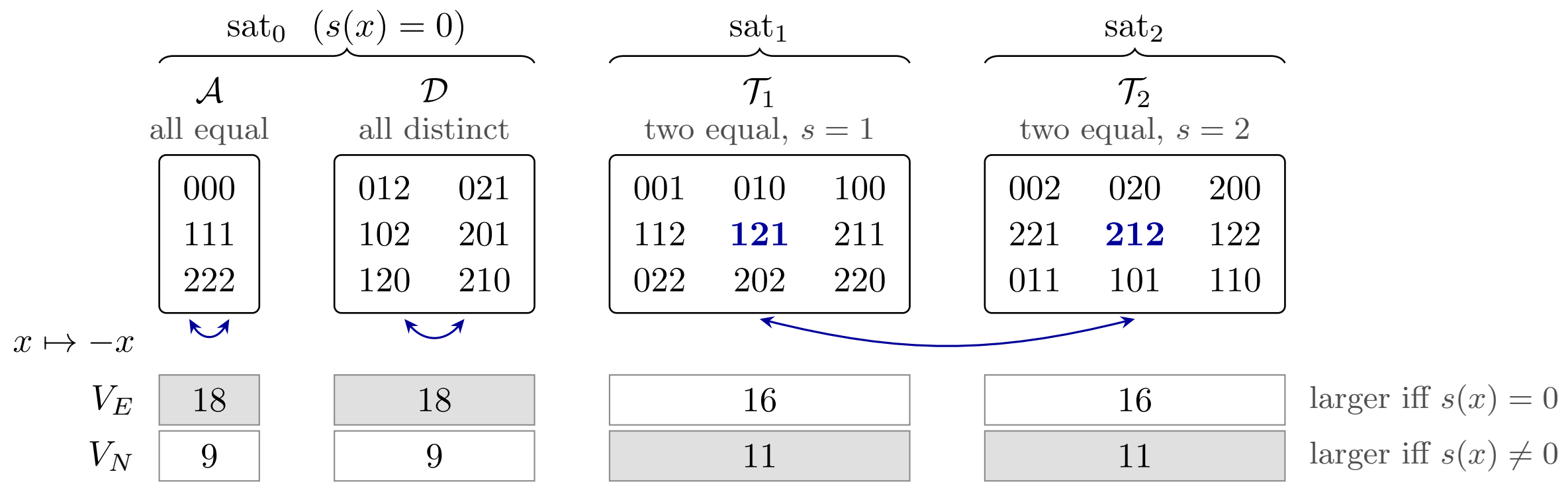

Here $k=3$, and the gadgets do not use their global vertices at all. Write $s(x)=x_1+x_2+x_3\in\Z_3$
and observe that $s(x)=0$ iff the three colours are all equal or all distinct, and $s(x)\neq 0$ iff
exactly two of them are equal (\cref{fig:tuples}).

\begin{itemize}
\item The \emph{E-gadget} $E$ (\cref{fig:gadgets}(a)) is the type-$0$ gadget of~\cite[Lemma~C.1]{NRT25}: auxiliary vertices
$a_1,\dots,a_6$ and the $18$ unit-weight edges
\[
\begin{array}{l}
p_1a_1,\ p_1a_2,\ p_1a_3,\ p_1a_4,\quad p_2a_3,\ p_2a_4,\ p_2a_5,\ p_2a_6,\quad
p_3a_1,\ p_3a_2,\ p_3a_5,\ p_3a_6,\\
a_1a_6,\ a_1a_4,\ a_2a_3,\ a_2a_5,\ a_3a_6,\ a_4a_5.
\end{array}
\]
The auxiliary vertices form the $6$-cycle $a_1a_4a_5a_2a_3a_6$, each is adjacent to two primaries, and
opposite vertices of the cycle are adjacent to the same pair of primaries; hence $E$ is invariant under all
permutations of the primaries.
\item The \emph{N-gadget} $N$ (\cref{fig:gadgets}) consists of a unit-weight triangle on $p_1,p_2,p_3$ and
one auxiliary vertex $q$ joined to $p_1,p_2,p_3$ by edges of weight $3$ (total weight $12$).
\end{itemize}

\begin{figure}[t]
\centering
\begin{tikzpicture}[x=1cm,y=1cm,
  prim/.style={circle,draw,fill=black!15,inner sep=0pt,minimum size=12pt,font=\footnotesize},
  aux/.style={circle,draw,fill=white,inner sep=0pt,minimum size=11pt,font=\scriptsize},
  univ/.style={rectangle,draw,fill=black!5,inner sep=0pt,minimum size=11pt,font=\scriptsize},
  e1/.style={line width=0.4pt},
  e1p/.style={line width=0.4pt,black!60},
  e3/.style={line width=1.2pt},
  u1/.style={line width=0.4pt,blue!60!black},
  u3/.style={line width=1.2pt,blue!60!black},
  lab/.style={font=\small}]
\begin{scope}[shift={(0,0)}]
  \node[aux] (aA1) at (90:1.7) {$a_1$};
  \node[aux] (aA4) at (30:1.7) {$a_4$};
  \node[aux] (aA5) at (-30:1.7) {$a_5$};
  \node[aux] (aA2) at (-90:1.7) {$a_2$};
  \node[aux] (aA3) at (-150:1.7) {$a_3$};
  \node[aux] (aA6) at (150:1.7) {$a_6$};
  \node[prim] (pA1) at (90:0.765) {$p_1$};
  \node[prim] (pA2) at (-30:0.765) {$p_2$};
  \node[prim] (pA3) at (210:0.765) {$p_3$};
  \draw[e1] (aA1) -- (aA4);
  \draw[e1] (aA4) -- (aA5);
  \draw[e1] (aA5) -- (aA2);
  \draw[e1] (aA2) -- (aA3);
  \draw[e1] (aA3) -- (aA6);
  \draw[e1] (aA6) -- (aA1);
  \draw[e1p] (pA1) -- (aA1);
  \draw[e1p] (pA1) -- (aA2);
  \draw[e1p] (pA1) -- (aA3);
  \draw[e1p] (pA1) -- (aA4);
  \draw[e1p] (pA2) -- (aA3);
  \draw[e1p] (pA2) -- (aA4);
  \draw[e1p] (pA2) -- (aA5);
  \draw[e1p] (pA2) -- (aA6);
  \draw[e1p] (pA3) -- (aA1);
  \draw[e1p] (pA3) -- (aA2);
  \draw[e1p] (pA3) -- (aA5);
  \draw[e1p] (pA3) -- (aA6);
  \node[lab] at (0,-2.75) {(a) $E$, total weight $18$};
\end{scope}
\begin{scope}[shift={(6.3,0)}]
  \node[aux] (aC1) at (90:1.7) {$a_1$};
  \node[aux] (aC4) at (30:1.7) {$a_4$};
  \node[aux] (aC5) at (-30:1.7) {$a_5$};
  \node[aux] (aC2) at (-90:1.7) {$a_2$};
  \node[aux] (aC3) at (-150:1.7) {$a_3$};
  \node[aux] (aC6) at (150:1.7) {$a_6$};
  \node[prim] (pC1) at (90:0.765) {$p_1$};
  \node[prim] (pC2) at (-30:0.765) {$p_2$};
  \node[prim] (pC3) at (210:0.765) {$p_3$};
  \node[univ] (zC1) at (-3.1,0) {$z_1$};
  \node[univ] (zC2) at (3.1,0) {$z_2$};
  \draw[e1] (aC1) -- (aC4);
  \draw[e1] (aC4) -- (aC5);
  \draw[e1] (aC5) -- (aC2);
  \draw[e1] (aC2) -- (aC3);
  \draw[e1] (aC3) -- (aC6);
  \draw[e1] (aC6) -- (aC1);
  \draw[e1p] (pC1) -- (aC1);
  \draw[e1p] (pC1) -- (aC2);
  \draw[e1p] (pC1) -- (aC3);
  \draw[e1p] (pC1) -- (aC4);
  \draw[e1p] (pC2) -- (aC3);
  \draw[e1p] (pC2) -- (aC4);
  \draw[e1p] (pC2) -- (aC5);
  \draw[e1p] (pC2) -- (aC6);
  \draw[e1p] (pC3) -- (aC1);
  \draw[e1p] (pC3) -- (aC2);
  \draw[e1p] (pC3) -- (aC5);
  \draw[e1p] (pC3) -- (aC6);
  \draw[u1] (zC1) -- (aC6);
  \draw[u1] (zC1) -- (aC3);
  \draw[u1] (zC2) -- (aC4);
  \draw[u1] (zC2) -- (aC5);
  \begin{pgfinterruptboundingbox}
  \draw[u1] (zC1) .. controls (-2.43,0.74) and (-1.09,1.85) .. (aC1);
  \draw[u1] (zC1) .. controls (-2.43,-0.74) and (-1.09,-1.85) .. (aC2);
  \draw[u1] (zC1) .. controls (-2.47,2.52) and (0.37,3.21) .. (aC4);
  \draw[u1] (zC1) .. controls (-2.47,-2.52) and (0.37,-3.21) .. (aC5);
  \draw[u1] (zC2) .. controls (2.43,0.74) and (1.09,1.85) .. (aC1);
  \draw[u1] (zC2) .. controls (2.43,-0.74) and (1.09,-1.85) .. (aC2);
  \draw[u1] (zC2) .. controls (2.47,2.52) and (-0.37,3.21) .. (aC6);
  \draw[u1] (zC2) .. controls (2.47,-2.52) and (-0.37,-3.21) .. (aC3);
  \draw[u3] (zC1) .. controls (-3.71,3.47) and (3.71,3.47) .. (zC2);
  \end{pgfinterruptboundingbox}
  \path (-3.16,-2.36) rectangle (3.16,2.68);
  \node[lab] at (0,-2.75) {(c) $A_L$ for $L=2$, total weight $33$};
\end{scope}
\begin{scope}[shift={(0,-4.8)}]
  \node[prim] (pB1) at (90:1.2) {$p_1$};
  \node[prim] (pB2) at (-30:1.2) {$p_2$};
  \node[prim] (pB3) at (210:1.2) {$p_3$};
  \node[aux] (qB) at (0,0) {$q$};
  \draw[e1] (pB1) -- (pB2);
  \draw[e1] (pB2) -- (pB3);
  \draw[e1] (pB3) -- (pB1);
  \draw[e3] (qB) -- (pB1);
  \draw[e3] (qB) -- (pB2);
  \draw[e3] (qB) -- (pB3);
  \node[lab] at (0,-1.75) {(b) $N$, total weight $12$};
\end{scope}
\begin{scope}[shift={(6.3,-4.8)}]
  \node[prim] (pD1) at (90:1.2) {$p_1$};
  \node[prim] (pD2) at (-30:1.2) {$p_2$};
  \node[prim] (pD3) at (210:1.2) {$p_3$};
  \node[aux] (qD) at (0,0) {$q$};
  \node[univ] (zD1) at (-2.6,0.6) {$z_1$};
  \node[univ] (zD2) at (2.6,0.6) {$z_2$};
  \draw[e1] (pD1) -- (pD2);
  \draw[e1] (pD2) -- (pD3);
  \draw[e1] (pD3) -- (pD1);
  \draw[e3] (qD) -- (pD1);
  \draw[e3] (qD) -- (pD2);
  \draw[e3] (qD) -- (pD3);
  \draw[u3] (zD1) -- (qD);
  \draw[u3] (zD1) -- (pD1);
  \draw[u3] (zD1) -- (pD3);
  \draw[u3] (zD2) -- (qD);
  \draw[u3] (zD2) -- (pD1);
  \draw[u3] (zD2) -- (pD2);
  \begin{pgfinterruptboundingbox}
  \draw[u3] (zD1) .. controls (-2.4,-1.6) and (-0.15,-1.75) .. (pD2);
  \draw[u3] (zD2) .. controls (2.4,-1.6) and (0.15,-1.75) .. (pD3);
  \end{pgfinterruptboundingbox}
  \path (-2.63,-1.38) rectangle (2.63,0);
  \node[lab] at (0,-1.75) {(d) $N_L$ for $L=2$, total weight $36$};
\end{scope}
\end{tikzpicture}
\caption{Left: the gadgets of \cref{thm:main3}; right: their extensions for \MC{k} with $k=3+L$
(\cref{sec:family}), drawn for $L=2$. Thin edges have weight~$1$ and very thick edges weight~$3$; primaries
are grey, the universal vertices $z_1,\dots,z_L$ (the globals $g_3,\dots,g_{k-1}$) are squares, and edges at universal vertices are blue. (a)~The type-$0$
gadget $E$ of~\cite{NRT25}, whose auxiliary vertices form the $6$-cycle $a_1a_4a_5a_2a_3a_6$; (b)~the
gadget $N$ for types $1$ and~$2$; (c)~$A_L$ adds an edge of weight~$1$ from every $a_j$ to every $z_i$
(none from the primaries) and one of weight~$3$ between any two $z_i$; (d)~$N_L$ adds edges of weight~$3$
from every $z_i$ to $q$ and to every primary, and none between the $z_i$.}
\label{fig:gadgets}
\end{figure}

\begin{lemma}\label{lem:EN}
For all $x\in\Z_3^3$: $V_E(x)=18$ if $s(x)=0$ and $V_E(x)=16$ otherwise; $V_N(x)=11$ if $s(x)\neq 0$ and
$V_N(x)=9$ otherwise.
\end{lemma}

\begin{proof}
By symmetry under permutations of the primaries and of the colours, it suffices to consider
$x=(0,0,0)$, $x=(0,1,2)$ and $x=(0,0,1)$. For $(0,0,0)$, colour the cycle alternately with $1$ and $2$; for
$(0,1,2)$, give each $a_j$ the colour of the primary it is not adjacent to (consecutive vertices of the
cycle then differ). In both cases all $18$ edges are cut. For $x=(0,0,1)$, each of $a_1,a_2,a_5,a_6$ is
adjacent to a primary of colour $0$ and one of colour $1$, so it has an uncut edge unless it has colour
$2$; since $a_1a_6$ and $a_2a_5$ are disjoint edges, at least two edges are uncut. Colouring the cycle
$a_1a_4a_5a_2a_3a_6$ by $2,1,2,1,2,1$ leaves exactly
$p_3a_2$ and $p_3a_6$ uncut, so $V_E(x)=16$. (This agrees with~\cite[Lemma~C.1]{NRT25}, which states
completeness $18$ and value at most $16$ on unsatisfying tuples.)

For $N$: if exactly two of the $x_j$ are equal, the triangle contributes $2$ and $q$ can take the third
colour, contributing $9$. If all are equal, the triangle contributes $0$ and $q$ contributes $9$. If all
are distinct, the triangle contributes $3$ and $q$ agrees with one primary, contributing $6$.
\end{proof}

\begin{proof}[Proof of \cref{thm:main3}]
Use the family $(2E, N, N)$, i.e.\ $E$ with all weights doubled for type $0$ and $N$ for types $1,2$.
By \cref{lem:EN}, for every $x$ and every type $t$,
\begin{equation}\label{eq:identity}
V_t(x) \;=\; \beta_t + 2\big([s(x)=t]+[s(x)=-t]\big),\qquad \beta_0 = 32,\ \beta_1=\beta_2=9.
\end{equation}
Indeed, for $t=0$ the bracket is $2[s=0]$, and for $t\neq 0$ it is $[s\neq 0]$.
Let $\Phi$ be an instance from \cref{thm:hastad} and $J=J(\Phi)$, so $w(J)=\frac13(36+12+12)=20$.

\emph{Completeness:} by~\eqref{eq:completeness}, $\max_\chi\cut_\chi(J)\ge \frac13(36+11+11)-36\eps
= \frac{58}{3}-36\eps = (\frac{29}{30}-\frac{9}{5}\eps)\,w(J)$.

\emph{Soundness:} let $\chi$ be any colouring and $\sigma$ the assignment it induces on the variables of
$\Phi$. By~\eqref{eq:soundness-basic} and~\eqref{eq:identity},
\[
  \cut_\chi(J) \le \sum_e \lambda_e\beta_{t_e} + 2\sum_e \lambda_e\big([s_e=t_e]+[s_e=-t_e]\big)
  = \tfrac{50}{3} + 2\big(\val_\Phi(\sigma)+\val_\Phi(-\sigma)\big),
\]
where $s_e=\sigma(u_e)+\sigma(w_e)+\sigma(w'_e)$, because $-\sigma$ satisfies $e$ iff $-s_e=t_e$. If
$\val(\Phi)\le\frac13+\eps$, both values are at most $\frac13+\eps$, so
$\cut_\chi(J)\le \frac{50}{3}+\frac43+4\eps = 18+4\eps = (\frac{9}{10}+\frac{\eps}{5})\,w(J)$.
\end{proof}

\begin{remark}
The proof of \cref{thm:main3} uses neither role separation nor, essentially, balance: if the type-$0$
equations have total weight $\lambda_0$ (and types $1,2$ together $1-\lambda_0$), the same computation gives
the ratio $(23\lambda_0+\frac{31}{3})/(25\lambda_0+11)\le 31/33<16/17$. (For role-separated instances the
soundness hypothesis even forces balance in the NO case, see \cref{rem:balance}.)
\end{remark}

\begin{remark}
In the per-type framework of~\cite{NRT25,TSSW00}, $N$ is useless as a type-$1$ gadget: it does not
distinguish $s=1$ from $s=2$, so its value on unsatisfying inputs can equal its completeness. The gain
comes entirely from the analysis, which pairs $\sigma$ with $-\sigma$.
\end{remark}

\section{A soundness criterion for gadgets from H\aa stad's equations}\label{sec:S}

This section proves the criterion behind \cref{thm:main45,thm:barrier}. Readers interested only in
\MC{3} may skip it: \cref{thm:main3} was proved directly in \cref{sec:three}, and
\cref{sec:family,sec:barrier} use only the statement of \cref{thm:S}. Throughout, $k\ge 3$.

\begin{theorem}[Soundness criterion]\label{thm:S}
Let $(I_0,I_1,I_2)$ be a family of gadgets for \MC{k}, $k\ge 3$, such that each $V_t(x,y)$ is symmetric
under exchanging $x_2$ and $x_3$. Let
\[
  \rho = \frac{\max_{y\in\Y_k}\max_{x\in[k]^3}S_y(x)}{\sum_{t\in\Z_3}C_t}.
\]
Then for every $\delta>0$, \MC{k} is NP-hard to approximate within $\rho+\delta$. More precisely, with
$\bar w=\sum_t w(I_t)$, \MC{k} is $\big(\frac{\sum_t C_t}{\bar w}-\delta,\ \frac{\max_{x,y}S_y(x)}{\bar w}
+\delta\big)$-hard.
\end{theorem}

\paragraph{Proof idea.}
Fix a colouring $\chi$ of $J(\Phi)$, and let $P(t,x)$ be the weight of the type-$t$ equations whose three
variables receive the colours $x\in[k]^3$. By~\eqref{eq:soundness-basic}, the cut is at most
$\sum_{t,x}P(t,x)V_t(x,y)$. If, under $P$, the type $t$ were uniform and independent of the colour
pattern $x$, this would be an average of $\frac13S_y(x)$ over patterns, hence at most
$\frac13\max_xS_y(x)$. To compare $P$ with such a distribution, we decode $\chi$ into assignments of
$\Phi$: for any two maps $a,b\colon[k]\to\Z_3$, a $U$-variable of colour $c$ gets the value $a(c)$ and a
$W$-variable of colour $c$ gets the value $b(c)$ (\cref{fig:reduction}). In the NO case $\val(\Phi)\le\frac13+\eps$ of \cref{thm:hastad}, each
decoded assignment satisfies at most $\frac13+\eps$, and hence (shifting $a$) at least $\frac13-2\eps$, of
the equations, so every test $g_{a,b}$ defined below sees $t$ as nearly uniform and independent of $x$. The spanning lemma (\cref{lem:span}) shows that these tests
detect every dependence between $t$ and $x$ that can affect the cut.

Two features of the decoding explain the hypotheses. The two $W$-variables of an equation are decoded by
the same map $b$, so the tests cannot tell $x_2$ from $x_3$; this is why $V_t$ must be symmetric in them
(condition~(ii) of \cref{lem:span}). The $U$- and $W$-variables must be decoded by independent maps: the
proof of \cref{lem:span} shifts $a$ alone by a constant $c$, which changes the decoded sum by $c$. If $U$
and $W$ could overlap, only maps with $a=b$ would be available, a common shift would change the sum by
$3c=0$, and the argument would break down. This is why we need role-separated instances.

\paragraph{The tests.}
Let $\omega=e^{2\pi i/3}$. For maps $a,b\colon[k]\to\Z_3$ define $g_{a,b}\colon\Z_3\times[k]^3\to\{0,1\}$ by
\[
  g_{a,b}(t,x) = [\,a(x_1)+b(x_2)+b(x_3) = t\,].
\]
Thus $g_{a,b}(t,x)=1$ exactly when an equation of type $t$ whose variables $u_e,w_e,w'_e$ have the
colours $x_1,x_2,x_3$ is satisfied by the decoded assignment $u\mapsto a(\chi(u))$, $w\mapsto b(\chi(w))$.

\begin{lemma}[Spanning lemma]\label{lem:span}
Let $f\colon \Z_3\times[k]^3\to\R$ satisfy (i) $\sum_{t\in\Z_3} f(t,x)=0$ for every $x$, and (ii)
$f(t,x_1,x_2,x_3)=f(t,x_1,x_3,x_2)$. Then $f$ is a real linear combination of the functions
$g_{a,b}-\frac13$.
\end{lemma}

\begin{proof}
We expand $f$ in the characters $\omega^{\ell t}$ of the variable $t$. By~(i) the coefficient of
$\omega^{0}$ vanishes, and by~(ii) the coefficients of $\omega^{t}$ and $\omega^{2t}$ are functions of $x$
in $\C^k\otimes\Sym^2(\C^k)$. We show that the products $A(x_1)B(x_2)B(x_3)$ that arise from the $g_{a,b}$
span this space.

We work over $\C$. For $c\in\Z_3$ let $a+c$ denote the map $v\mapsto a(v)+c$. For fixed $(t,x)$,
$g_{a+c,b}(t,x)=1$ for exactly one $c$, namely $c=d:=t-a(x_1)-b(x_2)-b(x_3)$. Hence for
$\ell\in\{1,2\}$,
\[
 h_{\ell,a,b}:=\sum_{c\in\Z_3}\omega^{\ell c}\big(g_{a+c,b}-\tfrac13\big) = \omega^{\ell d}
 = \omega^{\ell t}\,\omega^{-\ell a(x_1)}\,\omega^{-\ell b(x_2)}\,\omega^{-\ell b(x_3)} .
\]
Let $\mathcal R\subseteq\C^k$ be the set of vectors whose entries are cube roots of unity. As $a$ and $b$
range over all maps, $\omega^{-\ell a(\cdot)}$ and $\omega^{-\ell b(\cdot)}$ range over all of $\mathcal R$
(for both $\ell=1,2$). Thus the span contains $\omega^{\ell t}A(x_1)B(x_2)B(x_3)$ for all $A,B\in\mathcal R$.

$\mathcal R$ spans $\C^k$: it contains $\mathbf 1$ and $\mathbf 1+(\omega-1)e_i$. The tensors $B\otimes B$,
$B\in\mathcal R$, span the symmetric square $\Sym^2(\C^k)$: if a symmetric matrix $Q$ satisfies
$\sum_{i,j}Q_{ij}B_iB_j=0$ for all $B\in\mathcal R$, then writing $B_i=\omega^{\theta_i}$ this says
$\sum_{i\le j}Q'_{ij}\,\omega^{\theta_i+\theta_j}=0$ for all $\theta\in\Z_3^k$, with
$Q'_{ii}=Q_{ii}$ and $Q'_{ij}=2Q_{ij}$ for $i<j$. The functions $\theta\mapsto\omega^{\theta_i+\theta_j}$
($i\le j$) are distinct characters of $\Z_3^k$ (their exponent vectors $e_i+e_j$ are distinct since
$2\not\equiv 0,1 \pmod 3$), hence linearly independent, so $Q=0$. Consequently the span contains
$\omega^{\ell t}\varphi(x_1)\psi(x_2,x_3)$ for every $\varphi\in\C^k$, every symmetric
$\psi\in\Sym^2(\C^k)$ and $\ell=1,2$.

Now expand $f(t,x)=\sum_{\ell\in\Z_3}\omega^{\ell t}\hat f_\ell(x)$ with
$\hat f_\ell(x)=\frac13\sum_t\omega^{-\ell t}f(t,x)$. Condition (i) gives $\hat f_0=0$, and (ii) makes
$\hat f_1,\hat f_2$ symmetric in $(x_2,x_3)$, i.e.\ elements of $\C^k\otimes\Sym^2(\C^k)$. So $f$ lies
in the complex span of the $g_{a,b}-\frac13$; since $f$ and the $g_{a,b}$ are real, taking real parts of
the coefficients gives a real combination.
\end{proof}

\begin{proof}[Proof of \cref{thm:S}]
Reduce from \cref{thm:hastad} as in \cref{sec:prelim}; then $w(J)=\bar w/3$. Completeness
is~\eqref{eq:completeness}. For soundness, fix $y\in\Y_k$. The function $f_y(t,x)=V_t(x,y)-\frac13 S_y(x)$
satisfies (i) and (ii) of \cref{lem:span}, so $f_y=\sum_{a,b}\gamma^y_{a,b}(g_{a,b}-\frac13)$ for some reals
$\gamma^y_{a,b}$. Let $K_\gamma=\max_{y}\sum_{a,b}|\gamma^y_{a,b}|$; it depends only on the family.

Let $\Phi$ satisfy $\val(\Phi)\le\frac13+\eps$, let $\chi$ be a colouring of $J(\Phi)$, and normalise so that
the globals have colours $y\in\Y_k$. Let $P(t,x)$ be the total weight of the equations $e$ with $t_e=t$
and $(\chi(u_e),\chi(w_e),\chi(w'_e))=x$, and let $\bar P(x)=\sum_tP(t,x)$; both are probability
distributions, since the weights $\lambda_e$ sum to $1$. For any maps $a,b\colon[k]\to\Z_3$, the decoded
assignment $\sigma_{a,b}(u)=a(\chi(u))$ for $u\in U$ and $\sigma_{a,b}(w)=b(\chi(w))$ for $w\in W$ is well
defined because $U\cap W=\emptyset$, and it satisfies exactly the weight $\langle P,g_{a,b}\rangle$. Hence
$\langle P,g_{a,b}\rangle\le\frac13+\eps$. Since $\sum_{c}g_{a+c,b}\equiv 1$, also
$\langle P,g_{a,b}\rangle = 1-\sum_{c\neq 0}\langle P,g_{a+c,b}\rangle \ge\frac13-2\eps$. Therefore
\begin{align*}
  \cut_\chi(J) &\le \sum_{t,x}P(t,x)\,V_t(x,y)
     &&\text{by \eqref{eq:soundness-basic},}\\
  &= \sum_x \bar P(x)\,\frac{S_y(x)}{3}+\sum_{a,b}\gamma^y_{a,b}\Big\langle P,\,g_{a,b}-\tfrac13\Big\rangle
     &&\text{as $V_t=\tfrac13S_y+f_y$, expanding $f_y$,}\\
  &= \sum_x \bar P(x)\,\frac{S_y(x)}{3}+\sum_{a,b}\gamma^y_{a,b}\Big(\langle P,g_{a,b}\rangle-\tfrac13\Big)
     &&\text{as $\textstyle\sum_{t,x}P(t,x)=1$,}\\
  &\le \frac13\max_x S_y(x) + 2K_\gamma\eps
     &&\text{as $\langle P,g_{a,b}\rangle\in[\tfrac13-2\eps,\tfrac13+\eps]$.}
\end{align*}
Choosing $\eps$ small enough compared to $\delta$ finishes the proof.
\end{proof}

\paragraph{Tightness.}
The numerator of $\rho$ cannot be improved by any analysis that uses H\aa stad's theorem as a black box.

\begin{proposition}\label{prop:tight}
For every family, every $x^\ast\in[k]^3$ and every $y^\ast\in[k]^k$, there is a balanced role-separated
instance $\Phi_0$ with $\val_{\Phi_0}(\sigma)=\frac13$ for every assignment $\sigma$, and a colouring of
$J(\Phi_0)$ of value $\frac13S_{y^\ast}(x^\ast)$. Hence no soundness analysis that uses only the NO
conclusion of \cref{thm:hastad} can improve on the numerator of $\rho$.
\end{proposition}

\begin{proof}
Let $\Phi_0$ consist of one variable $u\in U$, two variables $w,w'\in W$ and the three equations
$u+w+w'=t$, $t\in\Z_3$, of weight $\frac13$ each (or a disjoint union of copies of it). Every assignment
satisfies exactly one of the three equations. Colour $u,w,w'$ by $x^\ast_1,x^\ast_2,x^\ast_3$, the globals by
$y^\ast$, and the auxiliary vertices optimally.
\end{proof}

In the language of the proof of \cref{thm:S}, $\Phi_0$ realises the distribution
$P(t,x)=\frac13[x=x^\ast]$, which satisfies $\langle P,g_{a,b}\rangle=\frac13$ for all $a,b$ and so passes
every test. Moreover, if the maximum in $\rho$ is attained at some $(x^\ast,y^\ast)$ with $x^\ast_2=x^\ast_3$, then
\emph{every} $J(\Phi)$ with $\Phi$ balanced has a colouring of value $\frac13\max_{x,y}S_y(x)$: colour all
of $U$ by $x^\ast_1$, all of $W$ by $x^\ast_2$ and the globals by $y^\ast$. The analysis is then exact for
every instance, not only for $\Phi_0$. This is the case for all families in this paper, where the maximum
is attained at a constant tuple.

\paragraph{Further properties.}
The criterion is consistent with the known algorithms: for every family, $\rho\ge\alpha_k$, where
$\alpha_k$ is the approximation ratio of the Frieze--Jerrum rounding~\cite{FJ97,dKPW04}
(\cref{prop:alg}). \Cref{app:remarks} also shows that the symmetry hypothesis of \cref{thm:S} costs nothing
(\cref{rem:sym}), that the theorem does not need balance in the NO case (\cref{rem:balance}), and that for
$k=3$ it is never worse than the per-type analysis of~\cite{NRT25} (\cref{rem:pertype}).

\section{A uniform family for Max-(3+L)-Cut}\label{sec:family}

Let $L\ge 0$ and $k=3+L$. Call the global vertices $g_3,\dots,g_{k-1}$ (colours $3,\dots,k-1$ in $\yc$)
the \emph{universal} vertices $z_1,\dots,z_L$; the globals $g_0,g_1,g_2$ are not used (\cref{fig:gadgets}(c),(d)).

\begin{itemize}
\item $A_L$: the E-gadget of \cref{sec:three} (unit weights), plus an edge of weight $1$ between every
auxiliary vertex $a_j$ and every universal vertex, plus an edge of weight $3$ between every pair of
universal vertices. Total weight $18+6L+3\binom L2$.
\item $N_L$: a unit-weight triangle on the primaries, one auxiliary vertex $q$ joined with weight $3$ to
$p_1,p_2,p_3$ and to every universal vertex, and an edge of weight $3$ between every primary and every
universal vertex. Total weight $12+12L$.
\end{itemize}
The family is $(2A_L, N_L, N_L)$. For $L=0$ it is the family of \cref{sec:three}.

\begin{lemma}\label{lem:family-values}
For every $L\ge0$, the gadgets $A_L$ and $N_L$ are invariant under all permutations of the primaries,
and the family $(2A_L,N_L,N_L)$ satisfies
\[
C_0=w(2A_L),\qquad C_1=C_2=w(N_L)-1,\qquad \max_{x,y}S_y(x)=\textstyle\sum_tC_t-4 .
\]
Consequently $\sum_tC_t=2\Gamma_L$ and $\sum_tw(I_t)=2\Gamma_L+2$, with $\Gamma_L=29+18L+3\binom L2$.
\end{lemma}

\begin{proof}
For a colouring of a gadget, write $\col(v)$ for the colour of a vertex $v$.

\emph{Completeness.} Let $x\in\{0,1,2\}^3$ and let the globals have colours $\yc$, so the universal
vertices have distinct colours $3,\dots,k-1$, different from all colours of the primaries. For $t=0$,
colour the $a_j$ as in the proof of \cref{lem:EN} (colours in $\{0,1,2\}$): all edges of $E$, all edges
$a_jz_i$ and all edges $z_iz_{i'}$ are cut, so $C_0=w(2A_L)$. For $t\neq0$, exactly two primaries are
equal; colour $q$ by the third colour of $\{0,1,2\}$: every edge except one triangle edge is cut, so
$C_1=C_2=w(N_L)-1$.

\emph{Soundness.} Fix $x\in[k]^3$ and $y$, and let $\ell_A=w(A_L)-V_{A_L}(x,y)$ and
$\ell_N=w(N_L)-V_{N_L}(x,y)$ be the least uncut weights. Since
$S_y(x)=2V_{A_L}+2V_{N_L}=\sum_tw(I_t)-2(\ell_A+\ell_N)=\sum_tC_t+2-2(\ell_A+\ell_N)$, it suffices to show
$\ell_A+\ell_N\ge3$.

In $N_L$ the vertex $q$ is optimally coloured with a colour that is least frequent among its neighbours,
so $\ell_N$ equals the uncut weight of the triangle, plus $3$ for every pair of a primary and a universal
vertex of the same colour, plus $3\min_cn_c$, where $n_c$ is the number of neighbours of $q$ (among
$p_1,p_2,p_3,z_1,\dots,z_L$) of colour $c$. In $A_L$ every pair
$z_iz_{i'}$ of equal colour costs $3$. Hence we may assume that no primary shares its colour with a
universal vertex, that the $z_i$ have distinct colours, and that some colour is missing among the
$k=3+L$ neighbours of $q$; otherwise $\ell_A\ge3$ or $\ell_N\ge3$. If $x$ is constant, the triangle gives
$\ell_N=3$. If $x_1,x_2,x_3$ are distinct, then the primaries and the $z_i$ use all $k$ colours, which we
excluded. If exactly two coincide, then by symmetry (permuting the primaries and the colours) $x=(0,0,1)$; the
triangle gives $\ell_N\ge1$, and exactly one colour $c^\ast$ lies outside
$\{0,1,\col(z_1),\dots,\col(z_L)\}$. Each of $a_1,a_2,a_5,a_6$ is adjacent to a primary of colour $0$, to a
primary of colour $1$, and to all $z_i$, so it has an uncut edge unless it has colour $c^\ast$. Since $a_1a_6$ and $a_2a_5$ are disjoint edges of $E$, each of the pairs
$\{a_1,a_6\}$ and $\{a_2,a_5\}$ contributes an uncut edge, so $\ell_A\ge2$. In all cases $\ell_A+\ell_N\ge3$.

Equality holds for constant $x$ and distinct universal colours different from $x_1$: then $\ell_N=3$, and
the $6$-cycle of $E$ can be coloured with the two colours outside $\{x_1,\col(z_1),\dots,\col(z_L)\}$, so
$\ell_A=0$. Finally $w(A_L)=18+6L+3\binom L2$ and $w(N_L)=12+12L$ give
$\sum_tC_t=2w(A_L)+2w(N_L)-2=2\Gamma_L$.
\end{proof}

\begin{proof}[Proof of \cref{thm:main45}]
By \cref{lem:family-values} and \cref{thm:S} with $\bar w=2\Gamma_L+2$, \MC{k} is
$\big(\frac{2\Gamma_L}{2\Gamma_L+2}-\delta,\frac{2\Gamma_L-4}{2\Gamma_L+2}+\delta\big)$-hard, and
$\rho=1-2/\Gamma_L$. For $L=0,1,2,3,4$, $\Gamma_L=29,47,68,92,119$.
\end{proof}

\begin{remark}\label{rem:larger-k}
We also confirmed \cref{lem:family-values} by computer, for $L\le 6$ (\cref{app:comp}, which also lists
the values for $L=1$). Placing the universal vertices inside the gadgets is more economical than adding
them to the final graph as in \cref{sec:universal}: combining \cref{thm:main3} with \cref{lem:universal}
gives only $47/49$ for \MC{4} and $35/36$ for \MC{5}.
\end{remark}

\begin{remark}[How the family was found]
\Cref{thm:S} makes the design of a family a single linear program (\cref{app:comp}). Seeded with $E$ and
$N$ augmented by universal vertices (the construction of \cref{sec:universal} applied to the gadgets of
\cref{sec:three}), this LP returned the ratios $45/47$ and $33/34$; we then simplified the optimal solutions
into the family above. Our column-generation runs for $k=4,5$ did not find better families.
\end{remark}

\section{Universal vertices and perfect completeness}\label{sec:universal}

The following reduction is due to Kann, Khanna, Lagergren and Panconesi~\cite[Sec.~3]{KKLP97} (for $L=1$)
and Guruswami and Sinop~\cite[Sec.~2.3]{GS13} (Appendix~D of the arXiv version; for $L\le 2$, where the base
graph is $K$-colourable with $3\mid K$ and the weights are $d_v/K$).

\begin{lemma}\label{lem:universal}
Let $G$ be a weighted graph with total weight $M$ and weighted degrees $d_v$, let $L\ge1$ and $\kappa\ge 0$.
Let $H$ be obtained by adding vertices $z_1,\dots,z_L$, joining each $z_i$ to every $v\in V(G)$ with
weight $d_v/3$, and joining every pair $z_i,z_j$ with weight $\kappa$. For a graph $G'$ let
$\mathrm{unc}_k(G')$ be the minimum weight of monochromatic edges over $k$-colourings of $G'$. Then
\[
  \min\{\kappa,\ \mathrm{unc}_3(G)\} \;\le\; \mathrm{unc}_{3+L}(H) \;\le\; \mathrm{unc}_3(G),
\]
and $w(H)=M(1+\frac{2L}3)+\binom L2\kappa$. If $G$ is $3$-colourable then $H$ is $(3+L)$-colourable.
\end{lemma}

\begin{proof}
The upper bound: give $z_1,\dots,z_L$ the colours $3,\dots,L+2$ and colour $G$ optimally with
$\{0,1,2\}$. For the lower bound, let $\chi$ be a $(3+L)$-colouring of $H$. If two $z_i$ share a colour,
$\chi$ loses at least $\kappa$. Otherwise the $z_i$ use $L$ distinct colours, and a vertex $v\in V(G)$ can
share its colour with at most one $z_i$. If it does, the edge $vz_i$ is monochromatic, of weight
$d_v/3$. Among the three colours not used by the $z_i$, some colour is used by neighbours of $v$ of total
weight at most $d_v/3$; recolouring $v$ with it does not increase the weight of monochromatic edges.
Repeating this for all such $v$ yields a colouring that uses only three colours on $G$, whose
monochromatic weight is at least $\mathrm{unc}_3(G)$. The total weight follows from $\sum_vd_v=2M$.
\end{proof}

If $G$ comes from a $(c,s)$-hard instance of \MC{3} with $\Delta_c=1-c$ and $\Delta_s=1-s$, choose
$\kappa=\Delta_sM$, so that $w(H)=m_LM$ with $m_L$ as in~\eqref{eq:universal-ratio}. Then
$\mathrm{unc}_{3+L}(H)\le\Delta_cM$ in the YES case and $\ge\Delta_sM$ in the NO case, so \MC{(3+L)} is
hard within
\begin{equation}\label{eq:universal-ratio}
  \frac{m_L-\Delta_s}{m_L-\Delta_c},\qquad m_L = 1+\tfrac{2L}{3}+\tbinom L2\Delta_s .
\end{equation}

\begin{proof}[Proof of \cref{thm:perfect}]
Apply~\eqref{eq:universal-ratio} to the $(1,\frac{16}{17}+\eps)$-hardness of $3$-colouring
of~\cite{AOTW14} ($\Delta_c=0$, $\Delta_s=\frac1{17}$): the ratio is
$1-\frac{\Delta_s}{m_L}=1-\frac{3}{51+34L+3\binom L2}$, e.g.\ $\frac{82}{85}$ for $L=1$ and $\frac{119}{122}$
for $L=2$. Perfect completeness is preserved.
\end{proof}

\begin{remark}
Where do the previously stated constants come from? Guruswami and Sinop first prove a bound for
$K$-colourable graphs, for every $K$ divisible by $3$, by a tensor-product construction, and then add
$k-K\le2$ universal vertices~\cite[Sec.~2.3]{GS13}. Their bound for $K$ colours loses a factor $K$ in the
gap compared with the bound for $3$ colours that it is built from, and they use it also for $K=3$, where
this loss is unnecessary; \cite{AOTW14} plugged their $16/17$ into the resulting formula. (For $k=3$ the
formula gives only $50/51$.) Starting from the $3$-colour bound directly, \eqref{eq:universal-ratio}
improves on the formula for every $k$ from $4$ to $20$; even the $32/33$ of~\cite{GS13} would give
$54/55$ for \MC{4}.
\end{remark}

\section{A barrier for gadgets from 3Lin(Z3) to Max-3-Cut}\label{sec:barrier}

Optimality proofs for gadgets via LP duality go back to~\cite{TSSW00}; \cite{HHMOW17} proved an analogous
optimality statement for gadgets from Chan's Hadamard predicates to $\textsc{2Lin}(\Z_2)$.

\begin{theorem}[computer-assisted]\label{thm:barrier}
For every family $(I_0,I_1,I_2)$ of \MC{3} gadgets (with any number of auxiliary vertices, possibly using
the global vertices),
\[
\max_{x\in\Z_3^3}S_{\yc}(x)\ \ge\ \tfrac{27}{29}\textstyle\sum_tC_t .
\]
Hence \cref{thm:S} cannot give a ratio below $27/29$ for \MC{3}, and the family of \cref{sec:three}
is optimal for it.
\end{theorem}

The proof exhibits a single distribution $\mu^\ast$ on $\Z_3^3$ such that, for every type $t$ and every
type-$t$ gadget, the average of $V_t(x,\yc)$ over $x\sim\mu^\ast$ is at least $\frac{27}{29}C_t$. Summing
over $t$ gives the theorem, since the maximum of $S_{\yc}$ is at least its average.

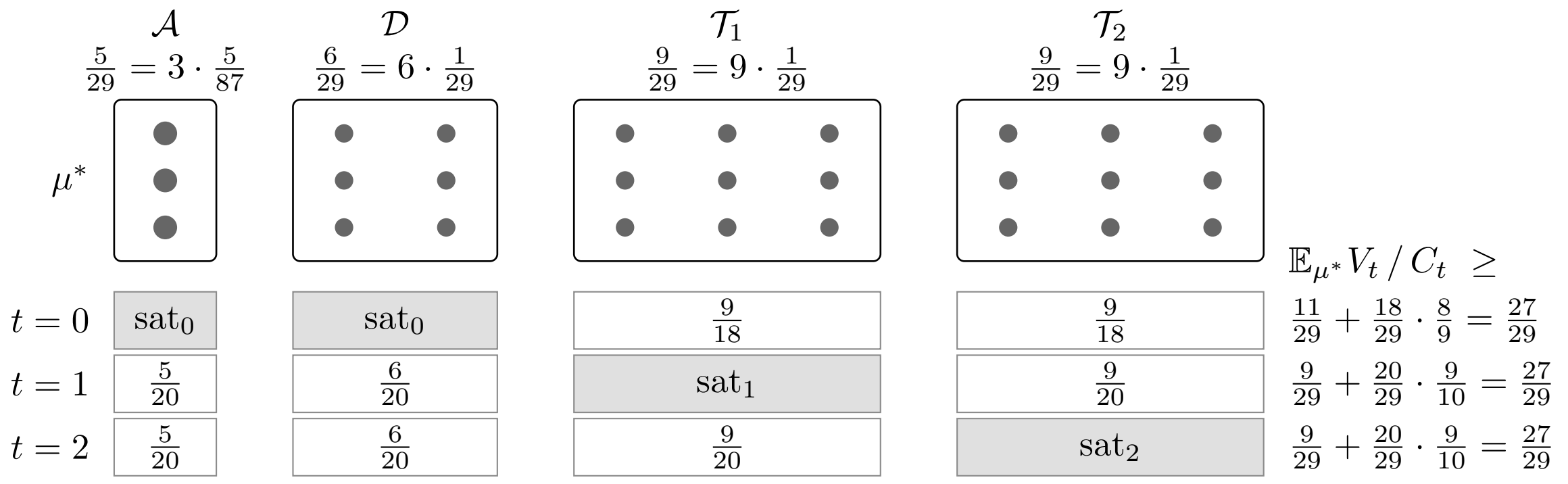
\begin{figure}[t]
\centering
\begin{tikzpicture}[x=1cm,y=1cm,
  cls/.style={draw,rounded corners=2pt,line width=0.5pt},
  cname/.style={font=\small,inner sep=1pt,text height=1.6ex,text depth=0.5ex},
  cmass/.style={font=\small,inner sep=1pt},
  rlab/.style={font=\small,anchor=east,inner sep=1pt},
  cell/.style={draw=black!45,line width=0.4pt},
  hi/.style={cell,fill=black!12},
  val/.style={font=\small},
  bnd/.style={font=\small,anchor=west,inner sep=1pt}]
\def\rh{0.46}\def\bt{1.58}
\def\ra{0.1162}
\def\rb{0.09}
\foreach \l/\r in {0/1, 1.75/3.75, 4.5/7.5, 8.25/11.25}
  \draw[cls] (\l,0) rectangle (\r,\bt);
\foreach \i in {0,1,2}{
  \fill[black!60] (0.5,\bt-0.1-\i*\rh-0.5*\rh) circle[radius=\ra];
  \foreach \x in {2.25,3.25,5,6,7,8.75,9.75,10.75}
    \fill[black!60] (\x,\bt-0.1-\i*\rh-0.5*\rh) circle[radius=\rb];}
\node[rlab] at (-0.2,0.5*\bt) {$\mu^\ast$};
\foreach \c/\n/\m in {0.5/{\mathcal A}/{\tfrac{5}{29}=3\cdot\tfrac{5}{87}},
                      2.75/{\mathcal D}/{\tfrac{6}{29}=6\cdot\tfrac{1}{29}},
                      6/{\mathcal T_1}/{\tfrac{9}{29}=9\cdot\tfrac{1}{29}},
                      9.75/{\mathcal T_2}/{\tfrac{9}{29}=9\cdot\tfrac{1}{29}}}{
  \node[cname] at (\c,\bt+0.72) {$\n$};
  \node[cmass] at (\c,\bt+0.3) {$\m$};}
\node[font=\small,anchor=south west,inner sep=1pt] at (11.45,-0.25) {$\mathbb{E}_{\mu^\ast}V_t\,/\,C_t\ \ge$};
\foreach \t/\a/\d/\p/\q/\bd in {
   0/sat/sat/{\tfrac{9}{18}}/{\tfrac{9}{18}}/{\tfrac{11}{29}+\tfrac{18}{29}\cdot\tfrac{8}{9}=\tfrac{27}{29}},
   1/{\tfrac{5}{20}}/{\tfrac{6}{20}}/sat/{\tfrac{9}{20}}/{\tfrac{9}{29}+\tfrac{20}{29}\cdot\tfrac{9}{10}=\tfrac{27}{29}},
   2/{\tfrac{5}{20}}/{\tfrac{6}{20}}/{\tfrac{9}{20}}/sat/{\tfrac{9}{29}+\tfrac{20}{29}\cdot\tfrac{9}{10}=\tfrac{27}{29}}}{
  \pgfmathsetmacro\yt{-0.3-\t*0.62}
  \pgfmathsetmacro\yb{\yt-0.56}
  \node[rlab] at (-0.2,{(\yt+\yb)/2}) {$t=\t$};
  \foreach \l/\r/\v in {0/1/\a, 1.75/3.75/\d, 4.5/7.5/\p, 8.25/11.25/\q}{
    \edef\tmp{\v}\def\satstr{sat}%
    \ifx\tmp\satstr
      \draw[hi] (\l,\yt) rectangle node[val] {$\sat_{\t}$} (\r,\yb);
    \else
      \draw[cell] (\l,\yt) rectangle node[val] {$\v$} (\r,\yb);
    \fi}
  \node[bnd] at (11.45,{(\yt+\yb)/2}) {$\bd$};}
\end{tikzpicture}
\caption{The distribution $\mu^\ast$ of the proof of \cref{thm:barrier}. Top: one dot per tuple, at the
positions of \cref{fig:tuples}, with area proportional to its mass ($\tfrac5{87}$ on each tuple of
$\mathcal A$, $\tfrac1{29}$ on every other tuple); the class masses are above the boxes. Row $t$: the
shaded cells form $\sat_t$, on which $V_t(x,\yc)\ge C_t$; each other cell gives the probability that $x_r$
lies in that class, for $r\sim\nu$ from the type-$t$ certificate of \cref{lem:dual}. These are exactly the
conditional probabilities of the classes under $\mu^\ast$, given that $x$ is unsatisfying for type $t$.
Right: the resulting lower bound on $\mathbb E_{\mu^\ast}V_t/C_t$, namely $\mu^\ast(\sat_t)$ plus the
unsatisfying mass times the certificate's guarantee $\sum_x\theta(x)$ (\cref{tab:certs}); it is
$\tfrac{27}{29}$ for every type.}
\label{fig:barrier}
\end{figure}

\paragraph{The distribution $\mu^\ast$.}
Partition $\Z_3^3$ into the classes $\mathcal{A}$ (all equal, $3$ tuples), $\mathcal{D}$ (all distinct, $6$
tuples), $\mathcal{T}_1$ and $\mathcal{T}_2$ (two equal with sum $1$, resp.\ $2$; $9$ tuples each), as in
\cref{fig:tuples}, so that $\sat_0=\mathcal A\cup\mathcal D$, $\sat_1=\mathcal T_1$ and
$\sat_2=\mathcal T_2$. Let $\mu^\ast$ be uniform within each class, with class masses in the proportions
\[
  \mu^\ast(\mathcal A):\mu^\ast(\mathcal D):\mu^\ast(\mathcal T_1):\mu^\ast(\mathcal T_2)=5:6:9:9,
\]
that is, mass $\frac5{29}$ on $\mathcal A$ and $\frac1{29}$ on every other tuple (\cref{fig:barrier}). A
type-$t$ gadget is guaranteed its completeness value $C_t$ on $\sat_t$. On the unsatisfying tuples we use a
certificate (\cref{lem:dual}) that guarantees on average a fraction $\frac89$ of $C_0$ for type $0$ and
$\frac9{10}$ of $C_t$ for $t=1,2$ (\cref{tab:certs}), but only for one particular distribution of inputs:
uniform on $\mathcal T_1\cup\mathcal T_2$ for type $0$, and in the proportions $5:6:9$ on
$\mathcal A,\mathcal D,\mathcal T_2$ for type $1$ (on $\mathcal A,\mathcal D,\mathcal T_1$ for type $2$).
These requirements determine $\mu^\ast$: it is the only distribution, uniform within each class, whose
conditional distribution on the unsatisfying tuples of each type is that of the certificate. With these
masses all three types end at the same lower bound on $\mathbb E_{\mu^\ast}V_t/C_t$, namely
$\mu^\ast(\sat_t)$ plus the unsatisfying mass times the certificate's fraction:
$\frac{11}{29}+\frac{18}{29}\cdot\frac89 =
\frac{9}{29}+\frac{20}{29}\cdot\frac{9}{10} = \frac{27}{29}$.

\paragraph{Certificates.}
The bounds on the unsatisfying tuples are instances of weak LP duality.

\begin{lemma}\label{lem:dual}
Fix $t$ and let $\mathcal F_t=[3]^{\sat_t}$ be the set of \emph{columns} (functions from
$\sat_t$ to colours). Suppose $\theta\colon\sat_t\to\R_{\ge0}$ and a probability
distribution $\nu$ over pairs $r=(x_r,\chi_r)$, with $x_r\in\Z_3^3$ and $\chi_r$ a colouring of the
\emph{universe} $\Omega=\{p_1,p_2,p_3,g_0,g_1,g_2\}\cup\mathcal F_t$ with $\chi_r(p_j)=(x_r)_j$ and
$\chi_r(g_i)=i$, satisfy, for every pair $\{\xi,\eta\}\subseteq\Omega$,
\begin{equation}\label{eq:dual-pair}
\Pr_{r\sim\nu}[\chi_r(\xi)\ne\chi_r(\eta)] \;\ge\; \sum_{x\in\sat_t}\theta(x)\,[\can_x(\xi)\ne\can_x(\eta)],
\end{equation}
where $\can_x$ is the canonical colouring of $\Omega$ at $x$: $\can_x(p_j)=x_j$, $\can_x(g_i)=i$,
$\can_x(f)=f(x)$. Then every type-$t$ gadget satisfies
$\mathbb E_{r\sim\nu}V_t(x_r,\yc)\ge\big(\sum_x\theta(x)\big)\,C_t$.
\end{lemma}

\begin{proof}
For each $x\in\sat_t$ fix a colouring of the auxiliary vertices of $I_t$ attaining $V_t(x,\yc)\ge C_t$.
Let $\phi\colon V(I_t)\to\Omega$ map $p_j$ and $g_i$ to themselves and each auxiliary vertex to its column
(its colour as a function of $x$). Then $\can_x\circ\phi$ is the fixed optimal colouring at $x$, and
$\chi_r\circ\phi$ is a colouring with primaries $x_r$ and globals $\yc$, of value at most $V_t(x_r,\yc)$.
Summing~\eqref{eq:dual-pair} over the edges $\xi'\eta'$ of $I_t$ with $\phi(\xi')\ne\phi(\eta')$, with their
non-negative weights, gives $\mathbb E_rV_t(x_r,\yc)\ge\sum_x\theta(x)V_t(x,\yc)\ge(\sum_x\theta(x))\,C_t$;
edges with $\phi(\xi')=\phi(\eta')$ are never cut by $\can_x\circ\phi$.
\end{proof}

\begin{table}[t]
\centering\small
\begin{tabular}{@{}cccccc@{}}
\toprule
type $t$ & $\sat_t$ & distribution of $x_r$ (by class) & $\sum_x\theta(x)$ & rows & common denominator\\
\midrule
$0$ & $\mathcal A\cup\mathcal D$ & uniform on $\mathcal T_1\cup\mathcal T_2$ & $8/9$ & $78$ & $324$\\
$1$ & $\mathcal T_1$ & $\mathcal A,\mathcal D,\mathcal T_2$ in proportions $5:6:9$ & $9/10$ & $162$ & $4680$\\
$2$ & $\mathcal T_2$ & $\mathcal A,\mathcal D,\mathcal T_1$ in proportions $5:6:9$ & $9/10$ & $162$ & $4680$\\
\bottomrule
\end{tabular}
\caption{The certificates of \cref{lem:dual} used for \cref{thm:barrier}. In each, $x_r$ is uniform within
each class and never satisfying; ``rows'' is the number of pairs $r$ (the \emph{rows} of the certificate) with positive weight under $\nu$, and the last column is
the common denominator of the stored exact certificate (\cref{app:comp}).}
\label{tab:certs}
\end{table}

\begin{proof}[Proof of \cref{thm:barrier}]
We use certificates as in \cref{lem:dual} with the parameters of \cref{tab:certs}. The type-$2$
certificate is obtained from the type-$1$ certificate by symmetry: for a type-$2$ gadget $I$, let $I'$ be
$I$ with $g_1$ and $g_2$ exchanged. By invariance under the colour permutation $c\mapsto-c$,
$V_{I'}(x,\yc)=V_I(-x,\yc)$, so $I'$ is a type-$1$ gadget with the same completeness, and the type-$1$
certificate gives $\mathbb E\,V_I(-x_r,\yc)\ge\frac9{10}C_I$, where $-x_r$ is distributed as in the last row
of \cref{tab:certs}. (We also stored and checked this mirrored certificate explicitly.)

In each certificate, the distribution of $x_r$ is the conditional distribution of $\mu^\ast$ on the
complement of $\sat_t$. Hence, since $V_t(x,\yc)\ge C_t$ on $\sat_t$,
\[
\mathbb E_{\mu^\ast}V_0\ge\tfrac{11}{29}C_0+\tfrac{18}{29}\cdot\tfrac89C_0=\tfrac{27}{29}C_0,\qquad
\mathbb E_{\mu^\ast}V_1\ge\tfrac{9}{29}C_1+\tfrac{20}{29}\cdot\tfrac9{10}C_1=\tfrac{27}{29}C_1,
\]
and likewise for $V_2$. Summing over $t$, $\max_xS_{\yc}(x)\ge\mathbb E_{\mu^\ast}S_{\yc}(x)
\ge\frac{27}{29}\sum_tC_t$.
\end{proof}

\begin{remark}
In each row $r$ of the certificates, the colour of a column $f$ is $h_r(f|_{R_r})$ for a set $R_r$ of one
or two satisfying tuples and a table $h_r\colon[3]^{R_r}\to[3]$. The certificates were found by solving
the dual LP with constraint generation over pairs, and then rounded to rationals (\cref{tab:certs}). The
rounded certificates are what is verified: \eqref{eq:dual-pair} was checked \emph{exactly}, for all pairs
of elements of $\Omega$ ($|\mathcal F_t|=3^9$), by two programs using different methods; the minimum slack
is $0$ (\cref{app:comp}).
\end{remark}

\section{Discussion}\label{sec:discussion}

\begin{itemize}
\item \emph{Perfect completeness.} Our reductions lose completeness ($29/30$); on $3$-colourable graphs
$16/17$~\cite{AOTW14} is still the best bound (the first constant bound there is due
to~\cite{Petrank94}). Can the negation trick be combined with a source that has perfect completeness?
\item \emph{Other sources.} By \cref{thm:barrier} and \cref{prop:tight}, per-type gadgets on the
instances of \cref{thm:hastad}, analysed through \cref{thm:hastad} alone, cannot beat $27/29$. Natural
candidates for other sources are the predicates $\sum_i a_ix_i=t$ over $\Z_3$ with $a\in\{\pm1\}^r$ and
$\sum_ia_i\equiv0$, which are NP-hard to approximate by Chan's theorem~\cite{Chan16} (the unconditional
analogue of~\cite{AM09}) and for which the negation trick again yields colour-symmetric predicates; gadget
reductions from Chan's predicates have been used successfully for $\textsc{2Lin}(\Z_2)$~\cite{HHMOW17}. We
expect the proof of \cref{thm:S} to carry over to Chan's $r$-partite instances, with one decoding map per
player (the analogue of \cref{lem:span} for products of $r$ arbitrary vectors), in which case no symmetry
hypothesis would be needed.
\item \emph{Optimality for $k\ge4$.} We do not know whether $45/47$ and $33/34$ are optimal under
\cref{thm:S}; our column generation found nothing better, but we have no barrier certificate.
\item \emph{Large $k$.} Our family gives $1-\Theta(1/k^2)$; combining \cref{thm:S} with the tensor
construction of~\cite{GS13} might improve the constant in $1-\Theta(1/k)$ and the range $k\le20$ in which
we improve the known bounds.
\end{itemize}

\appendix
\section{Further remarks on the soundness criterion}\label{app:remarks}

\begin{remark}[The symmetry hypothesis]\label{rem:sym}
The symmetry hypothesis of \cref{thm:S} is harmless: replace $I_t$ by
$I'_t=\frac12I_t\sqcup\frac12I_t^{\mathrm{sw}}$, where $I_t^{\mathrm{sw}}$ is $I_t$ with $p_2,p_3$ exchanged
and $\sqcup$ takes disjoint copies of the auxiliary vertices while sharing primaries and globals. Then
$V'_t(x,y)=\frac12(V_t(x,y)+V_t(x_1,x_3,x_2,y))$, so no $C_t$ decreases (the satisfying sets are symmetric)
and $\max_{x,y}S_y(x)$ does not increase. Gadgets that are symmetric under all permutations of the
primaries, as all gadgets in this paper, satisfy the hypothesis trivially.
\end{remark}

\begin{remark}[Balance]\label{rem:balance}
The proof of \cref{thm:S} does not use that the types are balanced in the NO case; with constant maps
$a\equiv c_1$, $b\equiv c_2$ the soundness hypothesis in fact forces every type to have weight
$\frac13\pm O(\eps)$.
\end{remark}

\begin{remark}[Comparison with the per-type analysis]\label{rem:pertype}
For $k=3$, \cref{thm:S} is never worse than the per-type analysis. Every $x\in\Z_3^3$ satisfies exactly
one type $t(x)$, so $S_y(x)\le V^+_{t(x)}+\sum_{t\ne t(x)}V^-_t$, where $V^+_t$ and $V^-_t$ are the
maxima of $V_t$ over satisfying resp.\ unsatisfying $x$ and all $y$; hence
$\rho\le(\sum_tV^-_t+\max_t(V^+_t-V^-_t))/\sum_tC_t$, which is the bound of the per-type reduction
theorem~\cite[Thm.~C.2]{NRT25}. With $V^+_t=C_t$ and $V^-_t=C_t-d$ this is $1-2d/\sum_tC_t$. \Cref{thm:S} can be much better because a tuple that is unsatisfying for two types need
not be bad for both gadgets simultaneously.
\end{remark}

\begin{proposition}\label{prop:alg}
For every family for \MC{k}, $\rho\ge\alpha_k$, where $\alpha_k$ is the approximation ratio of the
Frieze--Jerrum rounding~\cite{FJ97,dKPW04}.
\end{proposition}

\begin{proof}
Let
$\zeta_0,\dots,\zeta_{k-1}$ be unit vectors with $\langle \zeta_c,\zeta_{c'}\rangle=-\frac1{k-1}$ for
$c\ne c'$. For each $t$ and each satisfying tuple $x$ of type $t$ (there are $9$), fix a colouring
$\chi^t_x$ of $I_t$ with primaries $x$, globals $\yc$ and value $V_t(x,\yc)\ge C_t$, and let
$v^t_\xi=(9^{-1/2}\zeta_{\chi^t_x(\xi)})_x$ for each vertex $\xi$ of $I_t$. These are unit vectors with
$\langle v^t_\xi,v^t_{\xi'}\rangle=1-\frac{k}{k-1}\Pr_x[\chi^t_x(\xi)\ne\chi^t_x(\xi')]\ge-\frac1{k-1}$, and their
Frieze--Jerrum objective $\sum w_{\xi\xi'}\frac{k-1}{k}(1-\langle v^t_\xi,v^t_{\xi'}\rangle)$ equals
$\mathbb E_xV_t(x,\yc)\ge C_t$. On the set $B$ of primary and global vertices, the disagreement
probabilities do not depend on $t$ (any two primaries are independent and uniform on $\{0,1,2\}$), so the
Gram matrix of $(v^t_\xi)_{\xi\in B}$ is the same for all $t$. Fix vectors $(f_\xi)_{\xi\in B}$ in some $\R^m$ with this Gram
matrix, let $\Psi_t$ be the isometry from $\mathrm{span}\{v^t_\xi:\xi\in B\}$ onto $\mathrm{span}\{f_\xi\}$ with
$v^t_\xi\mapsto f_\xi$, and $\Pi_t$ the orthogonal projection onto the former span. In a direct sum
$\R^m\oplus H_0\oplus H_1\oplus H_2$ put $\tilde v_\xi=\Psi_t\Pi_tv^t_\xi+\iota_t(v^t_\xi-\Pi_tv^t_\xi)$ for $\xi$ in $I_t$,
where $\iota_t$ embeds into $H_t$. Then $\tilde v_\xi=f_\xi$ on $B$ and all inner products within each $I_t$ are
preserved, so this is a vector solution for $I_0\cup I_1\cup I_2$ (glued along $B$) with objective at least
$\sum_tC_t$, in which every edge has inner product $\ge-\frac1{k-1}$. (Auxiliary vertices of different
gadgets are not adjacent, so their inner products do not matter.) The rounding of~\cite{FJ97,dKPW04}
cuts each such edge with probability at least $\alpha_k\frac{k-1}{k}(1-\langle\tilde v_\xi,\tilde v_{\xi'}\rangle)$,
so some colouring has value at least $\alpha_k\sum_tC_t$; but every colouring has value at most
$\max_{x,y}S_y(x)$.
\end{proof}

\section{Methodology and computations}\label{app:comp}

\paragraph{How the results were obtained.}
The research was carried out with an AI system (Claude, by Anthropic) working in an agentic coding
environment, under the direction of the author, who set the goals at each stage (first improving the
gadgets of~\cite{NRT25} for \MC{4} and \MC{5}, then attacking \MC{3}) and decided which directions to
pursue. The AI system wrote and ran the linear-programming searches for gadgets and the verification
programs, proposed the negation trick of \cref{sec:three} while exploring alternative source problems,
derived the soundness criterion of \cref{sec:S}, found the gadget families of \cref{sec:family} (by the
linear program described below, followed by simplification) and the hand proof of
\cref{lem:family-values}, observed the consequences of the universal-vertex reduction in
\cref{sec:universal}, computed the certificates of \cref{sec:barrier}, searched the literature, and drafted
the text. Every main claim was checked in two ways. First, independently instantiated AI agents, which did
not see the reasoning that produced a claim, were asked to refute it: they re-read the primary sources
(in particular H\aa stad's verifier and the constants of~\cite{GS13,AOTW14}), re-derived the proofs, and
re-implemented the computations from scratch; four such referees also reviewed a draft of this paper.
Second, every gadget value and every certificate was verified by at least two independently written
programs, all using exact arithmetic, as described below.

\paragraph{Availability.}
The code, the gadget families and the certificates are available at
\url{https://github.com/ziangni-sys/max-3-cut}. All programs are in Python~3 with NumPy; HiGHS was used for the linear
programs.

\paragraph{Verification of families.} \texttt{scripts/verify\_sum\_criterion.py}, applied to the families
produced by \texttt{scripts/make\_unified\_family.py}, computes $V_t(x,y)$ for all $x\in[k]^3$ and
$y\in\Y_k$ by exhaustive enumeration of the auxiliary colourings in integer arithmetic, checks the
symmetry hypothesis of \cref{thm:S}, and outputs the exact ratio $\rho$; it reproduces
\cref{lem:family-values} for $L\le3$. The independently written \texttt{scripts/independent/family\_dp.py}
(which optimises the auxiliary vertices of $A_L$ by dynamic programming along the $6$-cycle) reproduces it
for $L\le6$. For $L=1$, \cref{tab:L1} lists the values by the pattern of equalities among
$x_1,x_2,x_3$ and $\col(z_1)$; the remaining patterns follow by permuting the primaries and the colours.

\begin{table}[h]
\centering\small
\begin{tabular}{@{}lccccccc@{}}
\toprule
$(x_1,x_2,x_3\,|\,\col(z_1))$ & $(0,0,0\,|\,0)$ & $(0,0,0\,|\,1)$ & $(0,0,1\,|\,0)$ & $(0,0,1\,|\,1)$
 & $(0,0,1\,|\,2)$ & $(0,1,2\,|\,0)$ & $(0,1,2\,|\,3)$\\
\midrule
$V_{2A_1}$ & $48$ & $48$ & $48$ & $48$ & $44$ & $48$ & $48$\\
$V_{N_1}$ & $12$ & $21$ & $17$ & $20$ & $23$ & $21$ & $21$\\
$S=V_{2A_1}+2V_{N_1}$ & $72$ & $90$ & $82$ & $88$ & $90$ & $90$ & $90$\\
\bottomrule
\end{tabular}
\caption{Gadget values of the family $(2A_1,N_1,N_1)$ for \MC{4} ($L=1$), by colour pattern.}
\label{tab:L1}
\end{table}
 \noindent\texttt{scripts/verify\_sym3.py} checks \cref{lem:EN} and the identity~\eqref{eq:identity};
the end-to-end reduction of \cref{sec:three} was also tested on random small instances of $\lin$ by brute
force.

\paragraph{The design LP.} For fixed candidate auxiliary vertices, each described by its column (its colour
in the completeness colouring at each satisfying tuple), each type-$t$ block has one variable per orbit of
edges under permutations of the primaries. Completeness rows require the canonical colouring at each
satisfying tuple to have value at least $c_t$; soundness is enforced by variables $v_t(x,y)$ with rows
$v_t(x,y)\ge$ (value of an explicit colouring), generated by local search and verified by exact
separation, and coupling rows $\tau\ge\sum_tv_t(x,y)$ for all $(x,y)$; we minimise $\tau$ subject to
$\sum_tc_t\ge1$. New auxiliary vertices were generated by exact dual pricing of single vertices (over all
$4^9$ columns for $k=4$) and by pricing whole small gadgets with up to $7$ auxiliary vertices; in our runs
the optimum stayed at $45/47$ ($k=4$) and $33/34$ ($k=5$).

\paragraph{Barrier certificates.} The certificates of \cref{thm:barrier} are stored as exact integers
(common denominator $D$, row weights, $\theta$, and for each row $x_r$, $R_r$, $h_r$) in the files
\texttt{cert\_t0.json}, \texttt{cert\_t1.json} and \texttt{cert\_t2.json} of the directory
\texttt{results/barrier\_k3}. The checker \texttt{check\_exact.py} verifies that
$\theta\ge0$, that $\nu$ is a probability distribution whose marginal on $x_r$ is uniform within each class,
that $\sum\theta$ equals $\frac89$ resp.\ $\frac9{10}$, and verifies~\eqref{eq:dual-pair} for all pairs of
elements of $\Omega$, including all $\binom{3^9}{2}$ pairs of columns, with integer-valued floating-point
matrix products whose partial sums stay below $2^{53}$ (hence exact). A second, independently written
checker (\texttt{check\_exact\_pairwise.py}) writes the slack of a pair of columns $(f,f')$ as a sum of
unary and pairwise terms in the nine pairs $(f(x),f'(x))$, $x\in\sat_t$, and minimises it exactly over
all $9^9$ choices of these pairs. Both find
minimum slack $0$.

\paragraph{Universal vertices.} \texttt{scripts/universal\_vertex.py} checks \cref{lem:universal} by brute
force on random weighted graphs and computes~\eqref{eq:universal-ratio}.

\section*{AI methodology}
\addcontentsline{toc}{section}{AI methodology}
We used Claude (Anthropic), a large language model, in an agentic coding environment (Claude Code,
including multi-agent workflows), to carry out this research and to write this paper. The tool proposed
the main ideas (the negation trick, the soundness criterion and its proof, the gadget families and the
proof of \cref{lem:family-values}, the use of the universal-vertex reduction, and the barrier
certificates), wrote and ran all search and verification programs, performed the literature search, and
drafted the text of all sections. It therefore materially affected all sections of the paper; the
methodology, including how the claims were checked by independent AI agents and by independent programs,
is described in \cref{app:comp}. The author directed the research and is responsible for the content of
the submission. The author verified the correctness and originality of all content, including references.

\bibliographystyle{alpha}
\bibliography{refs}

\newcommand{\etalchar}[1]{$^{#1}$}
\begin{thebibliography}{AOTW14}

\bibitem[AM09]{AM09}
Per Austrin and Elchanan Mossel.
\newblock Approximation resistant predicates from pairwise independence.
\newblock {\em Computational Complexity}, 18(2):249--271, 2009.

\bibitem[AOTW14]{AOTW14}
Per Austrin, Ryan O'Donnell, Li-Yang Tan, and John Wright.
\newblock New {NP}-hardness results for 3-coloring and 2-to-1 label cover.
\newblock {\em ACM Transactions on Computation Theory}, 6(1):Article 2, 2014.
\newblock arXiv:1210.5648.

\bibitem[Cha16]{Chan16}
Siu~On Chan.
\newblock Approximation resistance from pairwise-independent subgroups.
\newblock {\em Journal of the ACM}, 63(3):Article 27, 2016.
\newblock Preliminary version in STOC 2013.

\bibitem[CST01]{CST01}
Pierluigi Crescenzi, Riccardo Silvestri, and Luca Trevisan.
\newblock On weighted vs unweighted versions of combinatorial optimization
  problems.
\newblock {\em Information and Computation}, 167(1):10--26, 2001.

\bibitem[dKPW04]{dKPW04}
Etienne de~Klerk, Dmitrii~V. Pasechnik, and Joost~P. Warners.
\newblock On approximate graph colouring and {MAX}-$k$-{CUT} algorithms based
  on the $\vartheta$-function.
\newblock {\em Journal of Combinatorial Optimization}, 8(3):267--294, 2004.

\bibitem[FJ97]{FJ97}
Alan Frieze and Mark Jerrum.
\newblock Improved approximation algorithms for {MAX} $k$-{CUT} and {MAX}
  {BISECTION}.
\newblock {\em Algorithmica}, 18(1):67--81, 1997.

\bibitem[GS13]{GS13}
Venkatesan Guruswami and Ali~Kemal Sinop.
\newblock Improved inapproximability results for maximum $k$-colorable
  subgraph.
\newblock {\em Theory of Computing}, 9(11):413--435, 2013.
\newblock Preliminary version in APPROX 2009; arXiv:0910.2271.

\bibitem[GW04]{GW04}
Michel~X. Goemans and David~P. Williamson.
\newblock Approximation algorithms for {MAX-3-CUT} and other problems via
  complex semidefinite programming.
\newblock {\em Journal of Computer and System Sciences}, 68(2):442--470, 2004.
\newblock Preliminary version in STOC 2001.

\bibitem[H{\aa}s01]{Hastad01}
Johan H{\aa}stad.
\newblock Some optimal inapproximability results.
\newblock {\em Journal of the ACM}, 48(4):798--859, 2001.

\bibitem[Hei26]{Heilman26}
Steven Heilman.
\newblock Sharp hardness for {MAX-3-CUT} and quantum {MAX-CUT}.
\newblock arXiv:2608.00333, 2026.

\bibitem[HHM{\etalchar{+}}17]{HHMOW17}
Johan H{\aa}stad, Sangxia Huang, Rajsekar Manokaran, Ryan O'Donnell, and John
  Wright.
\newblock Improved {NP}-inapproximability for 2-variable linear equations.
\newblock {\em Theory of Computing}, 13(19):1--51, 2017.
\newblock Preliminary version in APPROX 2015.

\bibitem[IM12]{IM12}
Marcus Isaksson and Elchanan Mossel.
\newblock Maximally stable {G}aussian partitions with discrete applications.
\newblock {\em Israel Journal of Mathematics}, 189:347--396, 2012.

\bibitem[KKLP97]{KKLP97}
Viggo Kann, Sanjeev Khanna, Jens Lagergren, and Alessandro Panconesi.
\newblock On the hardness of approximating {Max} $k$-{Cut} and its dual.
\newblock {\em Chicago Journal of Theoretical Computer Science}, 1997:Article
  2, 1997.
\newblock Preliminary version in ISTCS 1996.

\bibitem[N{\etalchar{+}}25]{AlphaEvolve}
Alexander Novikov et~al.
\newblock {AlphaEvolve}: A coding agent for scientific and algorithmic
  discovery.
\newblock arXiv:2506.13131, 2025.

\bibitem[NRT25]{NRT25}
Ansh Nagda, Prabhakar Raghavan, and Abhradeep Thakurta.
\newblock Reinforced generation of combinatorial structures: Hardness of
  approximation.
\newblock arXiv:2509.18057, 2025.
\newblock Version 7, March 2026.

\bibitem[Pet94]{Petrank94}
Erez Petrank.
\newblock The hardness of approximation: gap location.
\newblock {\em Computational Complexity}, 4(2):133--157, 1994.

\bibitem[TSSW00]{TSSW00}
Luca Trevisan, Gregory~B. Sorkin, Madhu Sudan, and David~P. Williamson.
\newblock Gadgets, approximation, and linear programming.
\newblock {\em SIAM Journal on Computing}, 29(6):2074--2097, 2000.

\end{thebibliography}

\end{document}